\documentclass[useAMS,usenatbib]{mn2e}

\usepackage{epsfig}
\usepackage{myaasmacros}

\title[2-Dimensional Kinematics of Simulated Merger
  Remnants]{2-Dimensional Kinematics of Simulated Disc Merger Remnants}  
\author[Roland Jesseit, Thorsten Naab, Reynier Peletier and Andreas Burkert]
{Roland Jesseit$^{1}$\thanks{E-mail:
jesseit@usm.uni-muenchen.de;naab@usm.uni-muenchen.de; peletier@astro.rug.nl;andi@usm.uni-muenchen.de} , Thorsten
  Naab$^{1}$, Reynier Peletier$^{2}$  and Andreas Burkert$^{1}$%\footnotemark[1]\thanks{}% 
\\$^{1}$Universit\"ats Sternwarte M\"unchen, Scheinerstr.1, D-81679 M\"unchen, Germany 
\\$^{2}$  Kapteyn Astronomical Institute, University of Groningen, PO Box 800, 9700 AV Groningen, The Netherlands}
\begin{document}

\date{Draft version. Accepted ???. Received ??? in original form ???}
    
\pagerange{\pageref{firstpage}--\pageref{lastpage}} \pubyear{2002}

\maketitle
\label{firstpage}

\begin{abstract}
We present a two-dimensional kinematic analysis for a sample of 
simulated binary disc merger remnants with mass ratios 1:1 and 3:1. For the progenitor 
discs we used pure stellar models as well as models with 10\% of their mass in gas.
A multitude of phenomena also observed in real galaxies are found
in the simulations. These include misaligned rotation, embedded discs,
gas rings, counter-rotating cores and kinematic misaligned
discs. Using the 2D maps we illustrate projection effects
and the change in properties of a merger remnant when gas is included
in the merger. We find that kinematic peculiar subsystems are 
preferably formed in equal mass mergers. Equal-mass collisionless 
remnants  can show almost no rotation, regular rotation or strong 
kinematic misalignment. The inclusion of gas makes the remnants appear
more round(1:1) and axisymmetric(3:1). Counter-Rotating Cores (CRCs) are almost exclusively
formed in equal-mass mergers with a dissipational component. 3:1 remnants
show a much more regular structure. We quantify these properties by applying 
the kinemetric methods recently developed by Krajnovi\'c et al. This work 
will help to understand observations of elliptical galaxies with integral field spectrographs, 
like SAURON. \\ 
\end{abstract}


\section{Introduction}
N-body simulations of mergers of disk galaxies have been very succesful in reproducing  
many observational features of low and intermediate luminosity elliptical galaxies (\citealp{1972ApJ...178..623T};
\citealp{1992ApJ...393..484B}; \citealp{1992ApJ...400..460H}), while the most massive ellipticals are probably a 
result of collisionless early-type mergers or mixed mergers (\citealp{2003ApJ...597L.117K}; \citealp{2006ApJ...636L..81N}). 
However, the observational information obtained from  galaxies has been greatly enhanced 
with the advent of Integral Field Unit (IFU) instruments such as SAURON \citep{2001MNRAS.326...23B}, OASIS and GMOS. While 
mostly long-slit data was used to constrain the kinematic properties of early-type galaxies 
until recently, 2-dimensional data contains more information, and makes e.g. the detection
of a second distinct kinematic component, which might be aligned with an angle not covered by a slit in a 
conventional observation, much easier or even possible at all. 

The SAURON collaboration presented the first representative sample of 2-dimensional observations of Elliptical 
and S0 galaxies (\citealp{2004MNRAS.352..721E}, henceforth EM04). They presented 2D maps of line-of-sight 
velocities, dispersions and the third and fourth Gauss-Hermite moments. These maps showed a multitude of features 
such as kinematically decoupled components (KDCs, see also \citealp{2001ApJ...548L..33D}), central velocity dispersion 
dips, counter-rotating disks  and many others. Dynamical modelling of the kinematic data revealed that 
galaxies with significant rotation are more anisotropic than slow-rotating ellipticals \citep{2005nngu.confE...5C}. 
An interesting new correlation was found that the slow-rotating ellipticals host large KDCs which
have stellar populations of the same age as the main body, in contrast to fast-rotating ellipticals which have much smaller
KDCs and are younger on average \citep{2006astro.ph..9452M}. This correlation must contain some important information
on the formation history of early-type galaxies. 

Numerical studies of merging galaxies in the past showed that many peculiar kinematical features occur naturally
in merger remnants, especially when a dissipational component was present in the progenitor galaxies.
The dissipative component which can cool falls to the center and  changes the morpholgy of the remnant drastically 
by re-arranging its orbital content \citep{1996ApJ...471..115B}.
Disk like counter-rotating cores can also be formed from infalling gas, which has a spin vector with the opposite 
sign than the main rotating body \citep{1991Natur.354..210H}, however also dissipationless scenarios 
have been proposed for counter-rotating stellar populations in early-type 
galaxies (\citealp{1998ApJ...505L.109B};\citealp{1990ApJ...361..381B};
\citealp{2000MNRAS.316..315B}). \citet{2002MNRAS.333..481B} showed that the gaseous material which does not fall into 
the center can form large scale gas discs which can be warped or form bars. Polar-ring like features have been 
succesfully produced by several authors either in binary mergers \citep{1998ApJ...499..635B}, tidal accretion 
events \citep{2003A&A...401..817B} or in a cosmological context \citep{2006ApJ...636L..25M}. 

The aim of this work will be to compare 2-dimensional observations of a representative sample of 1:1 and 3:1 disc-disc 
merger remnants to the observations of the SAURON galaxy sample (EM04). The only similar study of this type was carried 
out by \citet{2000MNRAS.316..315B}, who identified peculiar  2-dimensional kinematic features in collisionless
merger models, such as as an orthogonally decoupled core, counter-rotating populations and misaligned rotation. Their sample contained
equal-mass mergers as well as mergers with mass ratios of 3:1 and they found that equal-mass mergers produce a larger variety 
of kinematical features. But as pointed out before, collisionless mergers are unlikely to be the only formation channel 
for early-type galaxies and it is unclear how the gas will influence 2D maps of such remnants. Gas infall also will vary from
merging geometry to merging geometry as the graviational torques on the gas will vary \citep{2002MNRAS.333..481B}. 

We will examine the merger sample of \citet{2006MNRAS.tmp..996N}, henceforth NJB06, which includes for every merging geometry 
a collisionless merger and a merger where each progenitor galaxy had 10\% of its disc mass in gas.
In total we study a sample of 96 mergers which should hopefully give us a similar variety of kinematic features than found in
the SAURON sample. As we can compare the mock obsevations of each collisionless remnant with its gaseous counterpart, we can 
easily assess which kinematical feature was already present in the dissipationless remnant and which was caused by the influence
of the gas..

Although 2-dimensional features can many times be intuitively grasped by visual inspection of the maps, this
is hardly a quantifiable way to compare to simulations. Recently \citet{2006MNRAS.366..787K}, henceforth K06,
developed a method to describe the properties of 2-dimensional maps of arbitrary moments of the line-of-sight-velocity distribution(LOSVD).
They termed this method {\it kinemetry} which works much in analogy to the common photometric fitting of isophotes.
With this method one can determine, for example, in an easy way position angle, amplitude and shape of the rotation of a galaxy, or merger 
remnant,  and identify kinematic subsystems more easily.
The shape of the 2-dimensional maps are of course viewing angle dependant, and we try to take this into account by analysing the projections
along the three principle axes. However, we are not carrying out a comprehensive statistical study in this work. We will use 
kinemetry throughout this work to quantify the kinematic features which arise through merging of disc galaxies. 
 
In Section \ref{sec:simu} we introduce the simulation sample. We explain the observational method to obtain 2D maps of
various moments of the LOSVD in Section \ref{sec:2dana}. The maps and their 
kinemetric analysis are presented in Section \ref{sec:results}. We compare these results
to 2D-kinematics of observed galaxies in Section \ref{sec:observ} and summarize our 
findings in Section \ref{sec:2dconcl}. Velocity maps and kinemetry of the whole sample 
are shown in the Appendix.

\section{Simulations}
\label{sec:simu}
The collisionless simulations are a subset of the simulations 
discussed in detail by \citet{2003ApJ...597..893N}, henceforth NB03. The simulations with gas, which we use for 
our analysis are identical to the ones presented in NJB06. In the following we give only 
the most important simulation parameters:

The progenitor disc galaxies were constructed in dynamical equilibrium using the
method described by \citet{1993ApJS...86..389H}. The system of units
was: gravitational constant G=1, exponential scale length of the
larger disc in the merger $h_d=1$ (the scale height was $h_z=0.2$) and
mass of the larger disc $M_d=1$. The discs were exponential with an
additional spherical, non-rotating bulge with mass $M_b = 1/3$, a
Hernquist density profile \citep{1990ApJ...356..359H} and a scale
length $r_b=0.2h$, and a pseudo-isothermal halo with a mass $M_d=5.8$,
cut-off radius $r_c=10h$ and collisionless core radius $\gamma=1h$. The parameters
for the individual components were the same as for the collisionless
mergers presented in NB03. 
For this study we have re-simulated the full set of 1:1 and 3:1 mergers 
with an additional gas component in the disc. We replaced 10\% of 
the stellar disc by gas with the same scale length and an initial  scale 
height of $h_{z,gas} = 0.1 h_z$. This is in agreement
with recent results from \citet{2006astro.ph..5436K} in which they show that
merging progenitor galaxies at low redshifts contain on average 10\% gas
for ellipticals more massive than $10^{10}$ solar masses.
The gas was represented by SPH particles assuming an isothermal equation of
state, $P = c_s^2\rho$, with a fixed sound speed of $c_s=0.039$ in
velocity units corresponding to $c_s \approx 10km/s$ if scaled to a
Milky Way type galaxy. The N-body/SPH simulations were performed 
using the hybrid N-body/SPH tree code VINE (Wetzstein et al. in prep.) 
with individual time steps. 

The galaxies approached each other on nearly parabolic orbits, in
agreement with predictions from cosmological simulations 
\citep{2006A&A...445..403K}, with an initial separation of 30 length units and a 
pericenter distance of 2 length units. The inclinations of the two discs
relative to the orbit plane were $i_1$ and $i_2$ with arguments of
pericenter $\omega_1$ and $ \omega_2$. In selecting unbiased initial
parameters for the disc inclinations we followed the procedure
described by \citet{1998giis.conf..275B}. The initial orientations for
the discs were the same as in NB03, Table
1. The merger remnants were allowed to settle into dynamical
equilibrium for approximately 10 dynamical timescales after the merger
was complete. Then their equilibrium state was analyzed.

The total sample comprises of 96 merger remnants of which half of them were
run with a gaseous component.
We adopt the following nomenclature for the mergers presented:
The first two numbers give the mass ratio of the progenitor discs, i.e.
11 or 31. The letter 'C' stands for analysing a collisionless merger (disc and bulge particles), 
'S' stabds for analysing only the stellar component of a gaseous merger (disc and bulge 
particles) and 'GS' analysing the total luminous part of a gaseous merger (disc, bulge and 
gas particles). The last number denotes the merging symmetry.
For example, the abbreviation 11C5 means: a collisionless equal mass 
merger with merging symmetry 5. The mergers we have analysed in detail, a comment
regarding their kinematic peculiarity and the detailed orbit parameters of the merger
they resulted from, can be found in Table \ref{tab:int}. 

\begin{table}
\begin{center}
\caption{Properties of the modeled merger remnants. Special kinematical features observed and
initial orientations of the progenitor discs are shown. \label{tab:int}}
\begin{tabular}{|l|l|c|c|c|c}
\hline \hline
Model       &  comment     & $i_1$ & $i_2$  & $\omega_1$ & $\omega_2$ \\
            &                                                         \\
\hline 
11C1        &  high $\sigma$ by counter-rotation    & 0    &   0  & 180   & 0           \\
11C3        &  regular rotation    & 0    &   0  & 71   & -30           \\
11C5        &  low rotation        & -109  &  -60   &  180   &     0     \\
11C8	    &  surface density change	 & -109  & -60    &	71   &    90 	\\
11C12       &  kinematic misalignment  & -109 & 0  &71  & 90  	\\
31C6        &  regular rotation       &  -109 & -60 & 71    &30      	\\
11S2	    &  low $\sigma$ ring    &0 & 0   &71   & 30 	\\
11S6        &  CRC in stars  &-109 &-60 & 71 &   30   	\\
11GS4       &  $\sigma$ double-peak   &-109 &-60 & 71 &   30   	\\
11GS9       &  polar ring 		&-109 &0 &180 &0   		\\
11GS16      &  CRC in gas   &-109 &60 &71 &90   		\\
31GS19      &  $\sigma$ dumbbell        &0 &0 &71 &-30  				\\

\hline 
\end{tabular}
\end{center}
\end{table}

\section{2-Dimensional Data Analysis}
\label{sec:2dana}
\subsection{LOSVDs}
Every remnant was projected along the long axis (X-axis, YZ-projection), 
the intermediate axis (Y-axis, XZ-projection) and short axis (Z-axis, XY-projection) of
the moment of inertia tensor defined by the 40\% most tightly bound
stellar particles. For the 2D analysis we 
binned particles within the central 3 length units on a
grid of $48 \times 48$ cells. This corresponds typically to 2-3 effective 
radii depending on projection (see \citealp{2006MNRAS.tmp..463N} for the exact 
determination of $r_e$). To include seeing effects we created
for every luminous particle $10\times 10$ pseudo-particles with
identical velocities on a regular grid  with a total size
of 0.125 unit lengths centered on the original particle
position. The mass of the original particle was then distributed to 
the pseudo-particles weighted by a Gaussian with a standard deviation 
of 0.1625 unit lengths. Thereafter the pseudo-particles were binned 
on a $48\times48$ grid.  

For the kinematic analysis we binned (mass weighted) all pseudo-particles 
falling within each grid cell in velocity along the line-of-sight. The width 
of the velocity bins was set to a value of 0.1 for line-of-sight velocities
$v_{\mathrm{los}}$ in the range $ -4 \le v_{\mathrm{los}} \le 4$. This
resulted in 80 velocity bins over the whole velocity interval. Using
the binned velocity data we constructed line-of-sight velocity
profiles (LOSVD) for each bin of the 2D
grid. Subsequently we parameterized deviations from the Gaussian shape of the velocity
profile using Gauss-Hermite basis functions
\citep{1993MNRAS.265..213G,1993ApJ...407..525V}. The kinematic
parameters of each profile ($\sigma_{\mathrm{fit}}$,
$v_{\mathrm{fit}}$, $H_3$, $H_4$) were then determined simultaneously 
by least squares fitting (\citealp{2001ApJ...554..291C, 2001ApJ...555L..91N}).

\subsection{Kinemetry}
Surface brightness and line-of-sight velocity are both moments of the stellar 
distribution function. The surface brightness is its zeroth-order and the line-of-sight 
velocity the first-order moment. However, there is a fundamental difference between 
the 2D fields of both moments: as the surface brightness is an even moment its iso-contours
are point-symmetric (and therefore closed) and as the line-of-sight-velocity is an odd
moment its iso-contours are point-{\it anti}-symmetric and therefore open. 
While isophotal shapes were found to be very close to perfect ellipses, such a choice is
not obvious for iso-velocity contours. K06 found empirically that the kinematics of early-type 
galaxies resemble very closely those of inclined discs, which in many cases can be well described
by circular motion. If you would project a circular orbit which moves in the plane of the galaxy on the sky,
then it would follow a simple cosine law: 
\begin{equation}
V(R,\psi)=V_0+V_C(R)\sin i \cos \psi,
\end{equation}
where R is the semi-major axis length of the projected circle on the 
sky, $V_0$ the systemic velocity, $V_C$ is the ring circular velocity,
$i$ is the ring inclination and $\psi$ is the azimuthal angle measured from 
the major axis of the ellipse (their Eq. 5).

However, as the authors state, elliptical galaxies are spheroids and not inclined discs 
and therefore their assumptions will be violated in nature. The next step is to 
see how far galaxies deviate from the simple cosine law. This can be achieved by the means   
of a Fourier series. A map of a moment of the LOSVD can be generalized to
\begin{equation}
\label{eq:map}
K(a,\psi)=A_0(a) + \sum_{n=1}^N A_n(a)\sin(n\psi)+B_n(a)\cos(n\psi),
\end{equation}
where $\psi$ is the eccentric anomaly and $a$ the semi-major axis of the 
ellipse (their Eq. 6). K06 find that only few terms are needed (up to third order) 
to find a good description of the 2D data.

In the case of an even moment the fitting procedure is similar to what is done
in conventional photometry, while in case of an odd moment their algorithm 
tries to minimize all terms of the harmonic expansion except $B_1(a)\cos(\psi)$, which
according to their Ansatz carries the major contribution to the map of an odd moment of the LOSVD. 
In the first stage of the fit, these terms are minimized on a grid of position angles ($\Gamma$) and 
flattenings ($q=1-\epsilon$). After the ellipse, which describes the velocity 
profile best, has been found, the rotation curve is extracted again and Fourier-analysed 
to arbritrary order.

We follow the authors in using the following four parameters to quantitatively 
describe kinematic 2D maps: 
The already mentioned $\Gamma$, which measures the alignment of the rotation and $q$ which can be 
interpreted as the opening angle of the iso-velocity contours.
They argue that although the amplitudes of the separate cosine and sine terms in principle 
describe different properties of the maps, it is better for systems which are not axisymmetric 
to collect the amplitudes of sine and cosine terms of the same order (their Eq. 10):
\begin{equation}
k_n=\sqrt{A_n^2 + B_n^2}.
\end{equation}
For n=1 we get the term $k_1$, which represents the amount of line-of-sight 
bulk rotation in case of a velocity map. The authors found that the ratio $k_5/k_1$, where $k_5$ is 
the first higher-order term of the expansion which has not been fitted to the map and expresses 
the deviations from regular rotation, is sensitive to the presence of multiple kinematic 
components in a galaxy. 

For a detailed description of their analysis method we refer the reader to the 
original paper K06.

\section{Results}
\label{sec:results}
We show in this section the most important 2-dimensional maps of our sample in order
to cover a broad range of differing kinematic features. The selection of these maps is again
guided by, but not restricted to, the results of EM04. We will use the results of our previous study on 
the orbital structure of merger remnants  (\citealp{2005MNRAS.360.1185J}, henceforth JNB05) and a kinemetric 
analysis for the interpretation of the maps. 
The apparent long axis of the remnants will be horizontally oriented for all the maps presented in this paper.
In general we show maps for the collisionless stellar particles (C), the stars of the simulations
with gas (S) and the stars in combination with gas (GS), as an indication of what the star would
look like in the case of late star formation. The newly 'formed' stars would  not have the
same stellar mass-to-light ratio than the 'old' stellar population. To account for this effect we adopt 
an ad hoc mass-to-light ratio following common stellar population models and modify the maps accordingly.
During the merger a significant fraction of the gas falls to the center (see NJB06) and overpowers the 
signal in the central bins and limits in some cases our interpretation of the results at very small radii.
All maps are shown in the original SAURON colour scheme as devised by Cappellari \& Emsellem (2001)
and included in the Kinemetry distribution.  
\subsection{Surface Density Maps}
\begin{figure*}
\begin{center}
  \epsfig{file=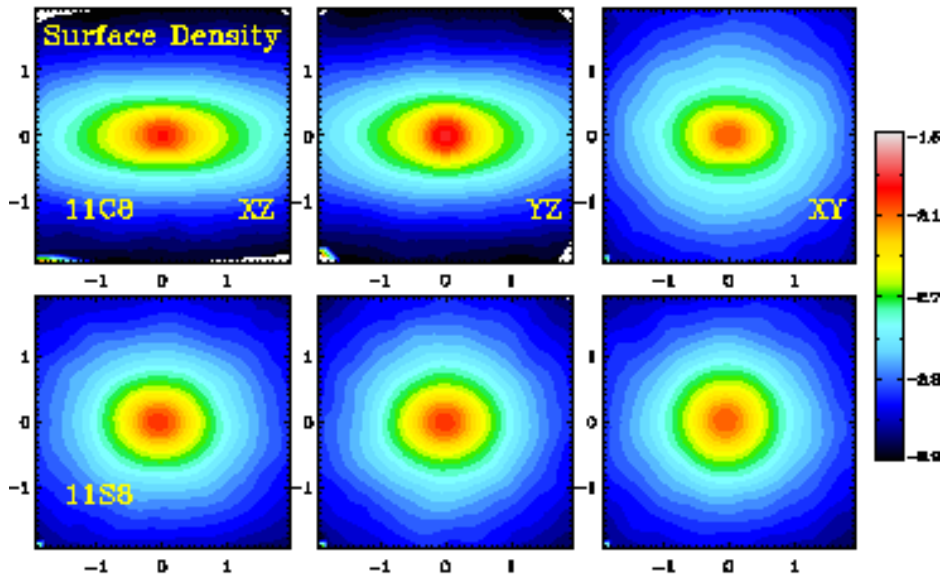, width=.7 \textwidth}
  \caption{2D surface density maps for the equal mass
    remnant with geometry 8. The projections along the three principal axes for the
    collisionless remnant (11C8, top row) are compared to the
    stellar component of the corresponding remnant with gas (11S8,
    bottom row). 11C8 is clearly triaxial whereas 11S8 is nearly round
    due to the influence of the additional gas component on the
    distribution of stars. One unit of lengths corresponds to 
    roughly one effective radius \citep{2006MNRAS.tmp..463N}. Colours indicate surface 
    density in logarithmic scale.} 
\label{fig:surfden1}
\end{center}
\end{figure*}

\begin{figure*}
\begin{center}
  \epsfig{file=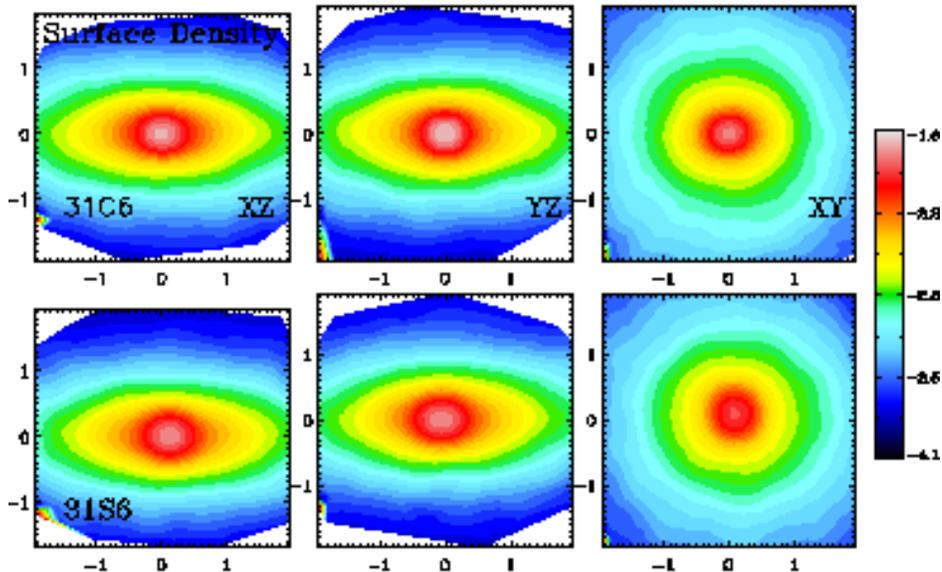, width=.7 \textwidth}
  \caption{Same as Fig.  \ref{fig:surfden1} for the 3:1 merger
  with geometry 6. There is no obvious influence of gas on the stellar distribution.}
\label{fig:surfden2}
\end{center}
\end{figure*}

It has been shown by a number of investigations that the presence of
gas during a merger event makes the remnant more axisymmetric
\citep{1996ApJ...471..115B}. NJB06 have shown that this effect is stronger 
for 1:1 remnants which are intrinsically more triaxial (or prolate) than 3:1 remnants. 
Fig. \ref{fig:surfden1} shows this effect for the 2D stellar surface density 
distribution of a collisionless 1:1 remnant and its counterpart, simulated with gas. Just by eye 
it is obvious that this particular remnant becomes much rounder if gas is included. Its 
intrinsic $b/a$ changes from 0.83 to 0.95, while $c/a$ 
increases even more significantly from 0.51 to 0.81. This is one 
of the few remnants in which the major axis tube fraction increases 
compared to the collisionless case from 5\% to 11\%. Next to minor axis tubes, 
also outer major axis tubes support potentials of spherical shape. So an increase of this fraction 
is to be expected (for a detailed description of orbit classification in merger remnants 
see JNB05). The effect of gas on the surface density maps of 3:1 remnants is 
in general hardly detectable (see Fig. \ref{fig:surfden2}) as the 
collisionless remnant is already rather axisymmetric. 

\subsection{Rotation of Collisionless Remnants}
\begin{figure*}
\begin{center}
  \epsfig{file=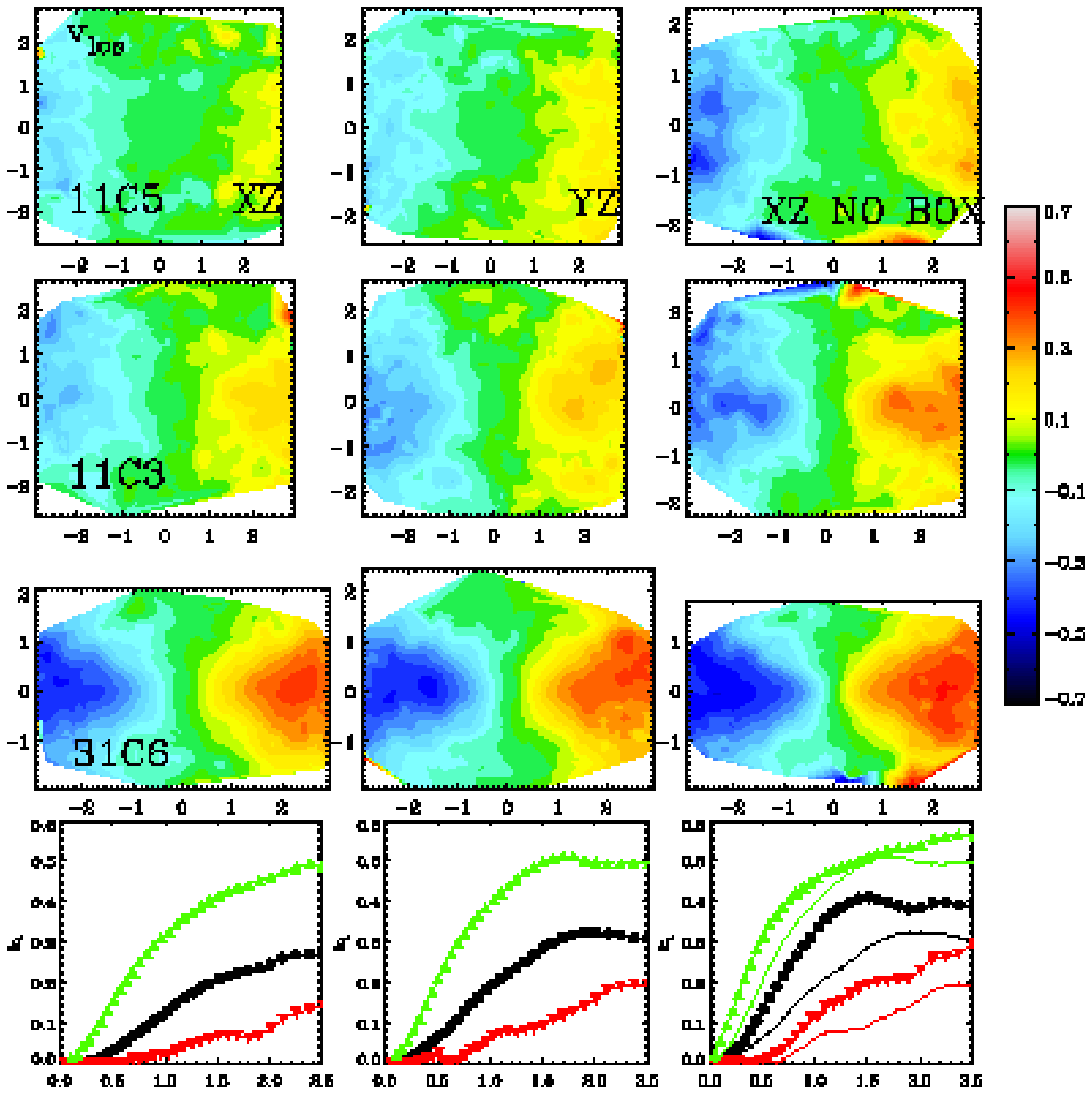, width=.8 \textwidth}
  \caption{2-dimensional line-of-sight velocity field for the
  collisionless 1:1 remnant 11C5 (top row), the 1:1 remnant 11C3 (second row) and 
  the 3:1 remnant 31C6 (third row). The bulk rotation is indicated in the bottom row. 31C6 (green triangles) and 11C3 (black circles)
  show regular rotation, while 11C5 (red upside  down triangles) has a large box orbit component and rotates very 
  little. How the 2-dimensional maps of the remnants would look like with the box orbits removed is shown in the last column.
   The rotation curve without box orbits is plotted with thick symbols and for comparison the original rotation curve with thinner symbols
   (bottom right). The net rotation increases significantly when the box orbits are removed. }
\label{fig:2drot}
\end{center}
\end{figure*}

In Fig. \ref{fig:2drot} we show the line-of-sight-velocity field of two collisionless 
1:1 remnants and one collisionless 3:1 remnant 31C6 with regular rotation. The properties of 3:1 
remnants do not change significantly with the initial disc orientations of the progenitors.
Therefore this example is typical for this mass ratio. The remnant 11C3 shows also regular 
rotation, despite of being an equal-mass merger, albeit with a lower amplitude than 31C6. 
11C3 is the 1:1 remnant with the highest minor axis tube fraction, but it is rather
an exception, as in general the rotational structure of equal-mass merger remnants 
is more complicated and depends on the initial disc geometry. 
In the most extreme case a 1:1 merger remnant can exhibit almost no rotation within in an 
effective radius, like 11C5, which has the highest box orbit fraction of all merger remnants. 

The kinemetry shows (Fig. \ref{fig:2drot}, bottom row), as indicated before, that 11C3 and 31C6 have a rising bulk rotation,
i.e. $k_1$ curve, while 11C5 has almost no amplitude in the center. We extract the box orbit 
population in each remnant, to see how this changes the maps. Mainly we observe now particles 
which have been classified as minor axis tubes, which have a significant amount of $L_z$.
In every case the amount of rotation increases (bottom row, right panel) and also the iso-velocity 
contours appear to be more closed. Still, all three maps are quite different in appearance. That means that in the 
merging process also the shape of the z-tubes is re-arranged. They are considerately more 'puffed up'
in equal-mass mergers than in un-equal mass mergers.

We want to note that the 2D velocity maps of 1:1 mergers are sometimes too complex to be fitted
adequately with ellipses. Very low rotation causes the ellipse fitting 
algorithm to extract erratic position angles. We therefore restrict ourselves ot constant position angle of $0^\circ$ 
and constant ellipticity of $q=0.5$. See also discussion in Sec. \ref{sec:misalign})

\subsection{The Influence of Gas on $v_{los}$ and $h_3$}
\label{sec:vh3}
\begin{figure*}
\begin{center}
  \epsfig{file=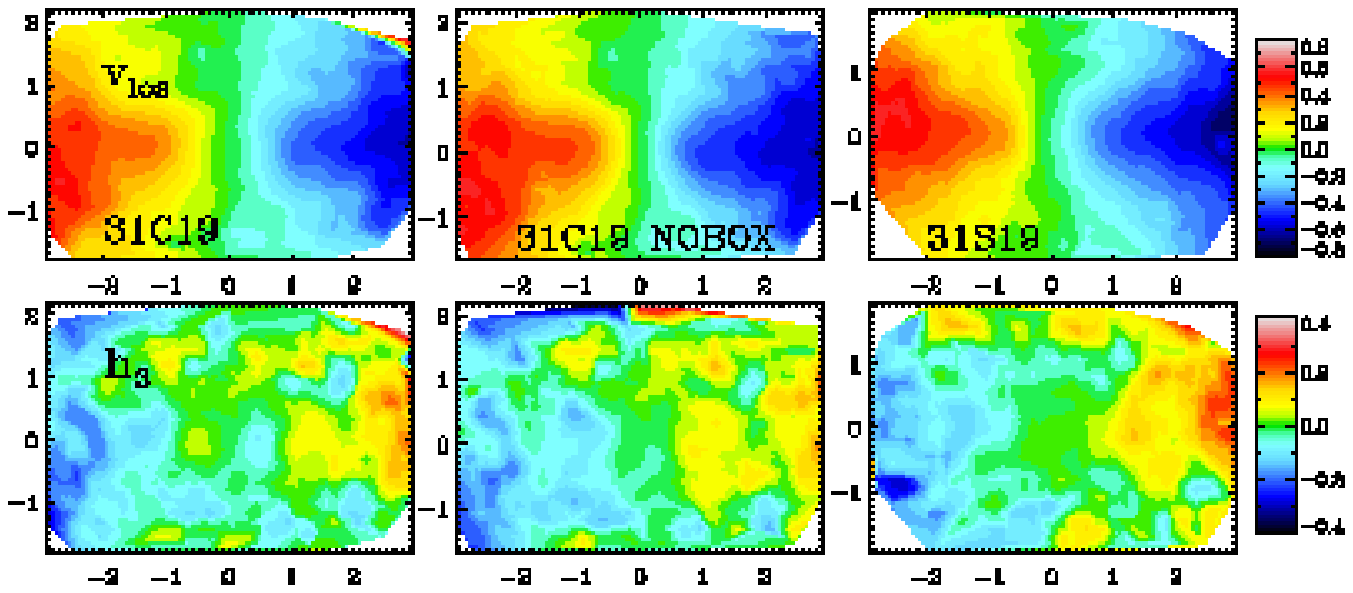, width=1. \textwidth}\\
   \epsfig{file=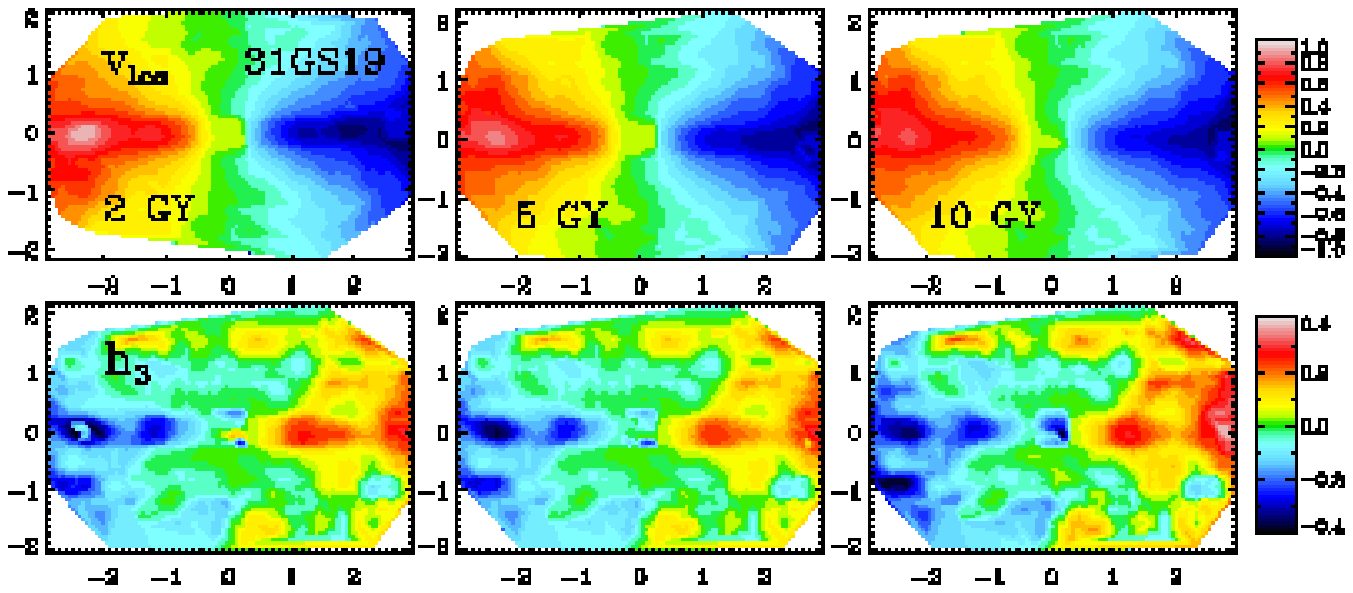, width=1. \textwidth}
  \caption{ 2d maps of velocity and $h_3$ for the 3:1 remnant with geometry 19. For the collisionless remnant (31C19)
  $v_\mathrm{los}$ and $h_3$ is correlated in the inner parts and
  anti-correlated in the outer parts (top left). This correlation vanishes, when the box orbits are removed (top middle).
Gas similarly changes the orbital setup and no correlation is visible in the stellar component of the gaseous remnant (top right).
  If the gas is 'ad hoc' turned into stars the  anti-correlation becomes even morevisible (two bottom rows). Accounting for the
   lower M/L of the gas particles increases the peak velocity but changes hardly the structure of the $h_3$ maps.}
\label{fig:2dvh3}
\end{center}
\end{figure*}

In Fig. \ref{fig:2dvh3} we show the 2D line-of-sight-velocity map of a 3:1 remnant and the corresponding
distribution of $h_3$. For the collisionless remnant (left panels)
$v_{\mathrm{los}}$ and $h_3$ are correlated inside 0.5 $r_e$ and anti-correlated in the outer parts. 
This behaviour is typical for collisionless remnants (\citealp{2000MNRAS.316..315B};
\citealp{2001ASPC..230..451N}; NJB06) and is not consistent with observations of elliptical 
galaxies \citep{1994MNRAS.269..785B}. 
The correlation of $v$ and $h_3$ at the center is caused by the superposition of box orbits and minor axis 
tubes in collisionless merger remnants. This happens because the peak of the total velocity profile at a radius close 
to the center is significantly moved to lower velocities, with respect to the short axis tube population, because box 
orbits, having zero mean angular momentum, peak normally at zero mean velocity. Therefore the normally steep leading wing 
becomes broader (see NJB06 for a more detailed discussion). When we remove the box orbit component
and observe this modified remnant, the correlation at the center has disappeared (Fig. \ref{fig:2dvh3}, top middle panel).

If gas is present during the merger the correlation at the center disappears even if only the
stellar component is considered (\ref{fig:2dvh3}, top row, right panel), leading to an overall
anti-correlation which is even stronger if we include the gas component in the analysis (two bottom rows). 
The gas influences the kinematic structure of the remnant two-fold: first, it is suppressing the population of 
box orbits \citep{1996ApJ...471..115B}, because the gradient of the potential in the center is very steep 
and orbits which come close to the center, i.e. box-like, are scattered on centrophobic, i.e. tube-like  orbits
\citep{1985MNRAS.216..467G}. Second, the gas particles themselves settle into a disk-like configuration. Both 
processes work towards a $v-h_3$ correlation more in agreement with what is found in real galaxies (NJB06).
 
As the light carried by particles on short axis tubes is important for the correct correlation, and as the gas 
particles move mostly in the XY-plane we want to test how a different stellar mass-to-light-ratio would
affect our results. We endorse a very simple method by giving more weight to the gas particles
relative to the stellar particles. To assign realistic M/L ratios we use the results of 
\citet{2003MNRAS.344.1000B} for a stellar population of solar metallicity Z=0.02, Salpeter IMF and in the K-Band. 
A 2 Gy old population of such stars would have a $M/L_K$, which is 2.527 lower and 5 Gy old population a  $M/L_K$
which is 1.336 times lower than a 10 Gy old population. The '10 Gy population' being the case when stars and gas particles
have the $M/L_k$, i.e. a normal 'GS' observation.
When we multiply the masses with the correction factors we see in Fig. \ref{fig:2dvh3} (two bottom rows) that 
the effect on the maps is minimal, but visible. The peak velocity in the 2 Gy observation is higher and the 
iso-velocity contours somewhat more closed, but our conclusions are unchanged.

\subsection{Kinematic Twists and Misalignment}
\label{sec:misalign}
\begin{figure*}
\begin{center}
  \epsfig{file=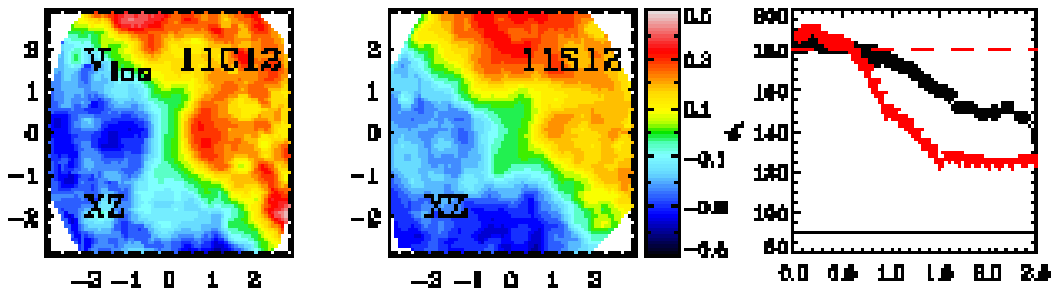, width=.85\textwidth}
  \epsfig{file=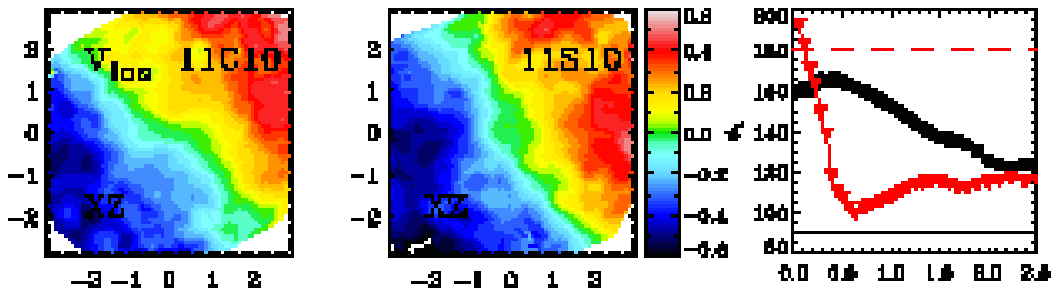, width=.85\textwidth}

  \caption{Velocity maps of two 1:1 merger remnants with geometry 10 and 12 of the stars of the collisionless and
gaseous remnants (column one and two). Kinematic position angle is shown in the third column for collisionless (black circles)
and gaseous (red triangles) remnant. Misaligned rotation ($180^\circ$) is indicated by a horizontal dashed red line and
aligned rotation by a continuos black line.  The maps of the collisionless remnants 
show strong kinematic misalignment (left column), which diminishes at larger radii. The inclusion of gas affects
both remnants differently. While the center of 11S10 exhibits a strong kinematic twist and then almost 
aligned rotation is the kinematic position angle of 11S12 hardly changed (last column). \label{fig:2dmisalign}}
\end{center}
\end{figure*}

A galaxy is kinematically misaligned if its rotation axis is not aligned with
its photometric major axis. Because of the strong kinematic misalignment the assumption of a constant 
position angle is of course wrong. To solve this problem we follow
the alternative procedure, proposed by K06, and extract 
the rotation curve by averaging over circles (see Appendix A of K06). Despite
this simplification the authors point out that $k_1$ and $k_5/k_1$ are still useful indicators 
for the bulk motion and deviations from regular motion. But because a circle has no 
defined position angle, we have to recover the phase angle

\begin{equation}
\phi_1=\arctan\left(\frac{A_1}{B_1}\right),
\end{equation}
where the phase angle $\phi_1$ represents the direction of the bulk rotation (their Eq. 10).

By convention signifies a $\phi_1=180^\circ$ misaligned rotation and $\phi_1=90^\circ$
alignment with the major photometric axis. As shown in the top row of Fig. \ref{fig:2dmisalign} 
the collisionless remnant 11C12 is at the center almost maximally misaligned with a $\phi_1 = 180^\circ$ and slowly changes 
to about $\phi_1=150^\circ$ at larger radii, i.e. the remnant has a kinematic twist (KT). 
This reflects the fact that the fraction of minor axis tubes increases with increasing radius, 
which skews the rotation more towards the major photometric axis. Nearly the same happens for the 
gaseous remnant (top row middle panel), only that the strongly misaligned region is smaller, as minor axis tubes are already
populated at smaller radii. The KT is slightly more pronounced, to about $\phi_1=125^\circ$,  between an outer disc-like 
component and a population of major axis tubes in the center.

In the second row of Fig. \ref{fig:2dmisalign} we show an example of 
another collisionless remnant, 11C10, with slightly lower misaligned rotation ranging from $\phi_1 = 160^\circ$ to
$125^\circ$. Interestingly, if re-simulated with gas (11S10), the stellar remnant 
shows a changed kinematic structure with respect to the collisionless remnant. The misalignment is almost 
completely removed. It is evident that the gas has to overcome a stronger
centrifugal barrier in the remnant 11S12  which partially prevents it to sink to the center and reorganize the orbital
structure of the remnant. In remnant 11S10 this is not true and the gas manages to accumulate in the XY plane re-arranging the
shape of the remnant. This is an indication that very strong misalignment can survive some degree of dissipation while 
small kinematical misalignment is wiped out more easily.

\subsection{The Flattening of Iso-Velocity Contours}
\label{sec:qparam}
K06 argued with the example of four early-type galaxies taken from
the SAURON sample that different kinematic flattenings, $q$, show to what extent a galaxy is an
axisymmetric  rotator, in which case $q$ 
remains constant with radius (see their Fig. 7). Two of their galaxies show peculiar 
flattenings, one with a radially rising $q$ and one with very wide opening angles
in the center (NGC 2549), but dropping to $q=0.4$ at larger radii. 

\begin{figure}
\begin{center}
  \epsfig{file=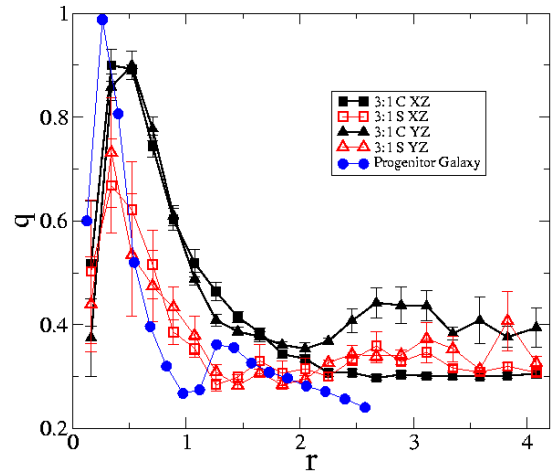, width=.85 \columnwidth}
  \caption{Flattening, $q$, of the iso-velocity contours vs radius. The $q$ parameter has been 
averaged at each radial bin for all 3:1 remnants for four cases: 3:1 collisionless merger XZ 
and YZ projection and 3:1 gaseous merger XZ and YZ-projection. Error bars indicate the rms at each bin. 
For comparison we plot the kinemetric fit of a progenitor disc galaxy}
\label{fig:kinemq}
\end{center}
\end{figure}
As the $q$ profile of NGC 2549 reminds us of flattenings found in 3:1 remnants, we will study in 
this section the influence of gas on the q-parameter in our 3:1 merger remnants. Their 
kinematics are much simpler than in 1:1 remnants, because it is much more difficult to populate major axis tube
orbits in a 3:1 merger remnant and hence their kinematic twists are almost negligible. 
Also, all of them show significant rotation in most of the projections, which is a precondition
to apply the kinemetric analysis. We analyze the YZ and the XZ projection, i.e. edge on projections,
of every remnant. To get an overview of how the $q$-parameter changes with projection and with the inclusion of 
gas, we calculate the mean $q$ for every radial bin and compare four subsamples: 3:1 collisionless YZ and XZ-projection 
and 3:1 dissipational mergers YZ and XZ-projection. We can see in Fig. \ref{fig:kinemq} 
that the iso-velocity contours of the XZ-projection of the collisionless remnants have in general a smaller opening 
angle in the outer regions than for the YZ-projection, reflecting the more extended, boxy shape of the minor 
axis tubes for the YZ projection (NJB06). The stellar remnants whose progenitors had a gaseous component, are almost 
axisymmetric and therefore the flattenings are not very different between the two projections. It is, however, evident 
from the plot that the $q$ of the stars in the gaseous remnants is significantly
lowered at radii smaller than two effective radii. Interestingly for large radii we find slightly
larger opening angles in the gaseous remnants as compared to the XZ-projection of the collisionless 
remnant. This is probably because at large radii the surviving part of the progenitor disc dominates
the rotation and the flattening of the initial disc is very low as indicated in 
Fig. \ref{fig:kinemq} (blue line).

\subsection{Systems with Polar Rings}
\label{sec:polar}
\begin{figure*}
\begin{center}
  \epsfig{file=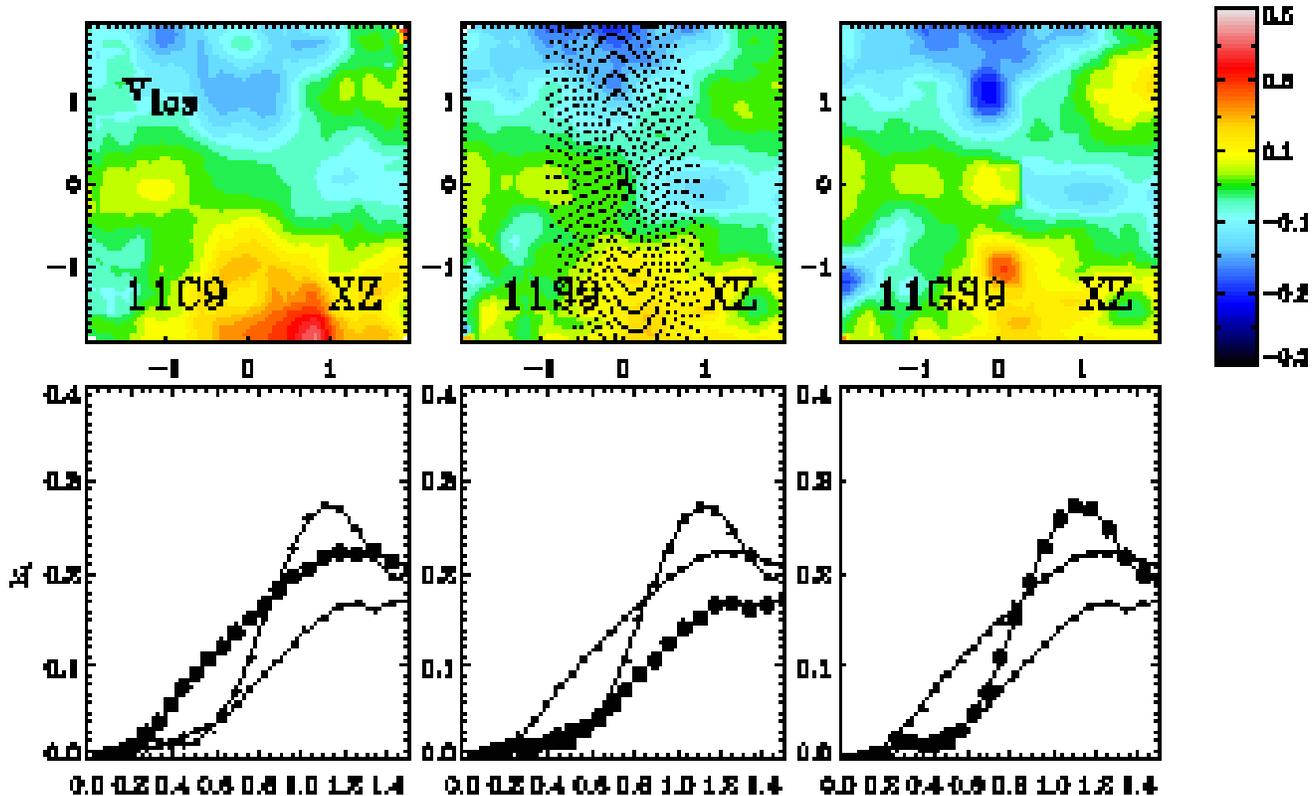, width=1. \textwidth}
  \caption{  {\it Top row:} Edge on projection (along the intermediate axis) of the 2D Velocity fields 
     of 11C9, 11S9 and 11GS9. The rotation curve was extracted along ellipses with $q=0.3$ and $\Gamma=95^\circ$ and are 
      overplotted in the top middle panel.. 
      Gas settles in a ring with a diameter of about two effective radii, inclined by $95^\circ$ with respect to the main stellar body. 
      The signature is similar to polar ring galaxies. Very little mass is contained in the polar gas ring 
      itself. {\it Bottom row:} The polar rotational structure is already present in the collisionless 
     remnant (bottom left panel, the bulk rotation of the corresponding map above is marked by thicker symbols). The 
     gas ring shows a higher rotation than the stellar component in the gaseous remnant.}
\label{fig:2dpolar}
\end{center}
\end{figure*}

A particularly interesting 1:1 remnant is shown in Fig. \ref{fig:2dpolar}. In 
this case the gas forms a large scale gas ring which is inclined by $\Gamma = 95^\circ$ with respect to the
photometric major axis of the galaxy and a central gaseous disc. The kinemetric
analysis reveals the different amount of rotation in the gas and in the stars. 
As the gas ring is very narrow we extract the rotation curve along an  
ellipse with a $q=0.3$ and with the same alignment, i.e. $\Gamma = 95^\circ$, as the ring.
As apparent from the bottom left panel of Fig. \ref{fig:2dpolar} already in the collisionless
remnant does a significant fraction of stars move on polar orbits. The gas also seems to have a higher
line-of-sight bulk motion than the stellar component. 

The polar ring is the result of the initial conditions of this particular merger for which
initially the discs of the progenitors are inclined to each other by nearly $90^\circ$. 
The properties of this remnant are similar to those of polar ring galaxies. \citet{1998ApJ...499..635B}
has shown that such systems can form in galaxy merging events. Our simulations
differ insofar that both progenitor discs have gas and that we ran a larger sample
of disc inclinations. Only in two out of 16 1:1 mergers was a significant 
amount of gas deposited in a polar ring. Typically this is just 10\% of
the initial gas mass or 1\% of the total baryonic mass of the system. The gas 
in such a polar ring stems from one of the merging partners only. We do not find any 
significant polar ring structures in 3:1 mergers.

Observed polar ring galaxies, though, have a comparable amount of mass in the polar 
ring and in the central component, typically an S0 galaxy (see e.g. 
\citealp{2000ASPC..197..119S}). Our simulations already
start with too little gas (10\% of the total mass) to form such a system. Alternatively, 
such systems can form by cold accretion \citep{2006ApJ...636L..25M} in a cosmological context.

\subsection{Counter-Rotating Cores}
\label{sec:crc}
The detection of Kinematically Decoupled Components (KDCs) at the center of early-type galaxies 
is an important indication for the hierarchical assembly of elliptical galaxies. A subset of these are central
stellar sub-systems, which rotate in the opposite direction to the outer stellar body, so called Counter-Rotating Cores (CRCs).
However, the exact process of how material, supposedly a left over from a merger, retains memory of its original angular 
momentum vector is a matter of debate. \citet{1984ApJ...287..577K} proposed that a dwarf satellite 
under certain pre-conditions can survive the tidal field of the  more massive progenitor and sink to 
the center. \citet{1990ApJ...361..381B} then showed in N-body simulations that such a formation scenario for a CRC
is viable for certain encounter geometries. It was demonstrated by \citet{1991Natur.354..210H} that dissipation can 
play an important role in forming central counter-rotating components (but see also \citealp{1998ApJ...505L.109B}). 
\citet{2002MNRAS.333..481B} elaborated on these findings, but pointed out that for the same encounter geometry
($i_1=71^\circ, \omega_1=90^\circ, i_2=-109^\circ, \omega_2=90^\circ$), but for a wider orbit no CRC is formed. The exact 
reason remains elusive. However, kinematically misaligned gaseous components were commonplace in their simulations, 
especially for retrograde encounters. 

\begin{figure*}
\begin{center}
  \epsfig{file=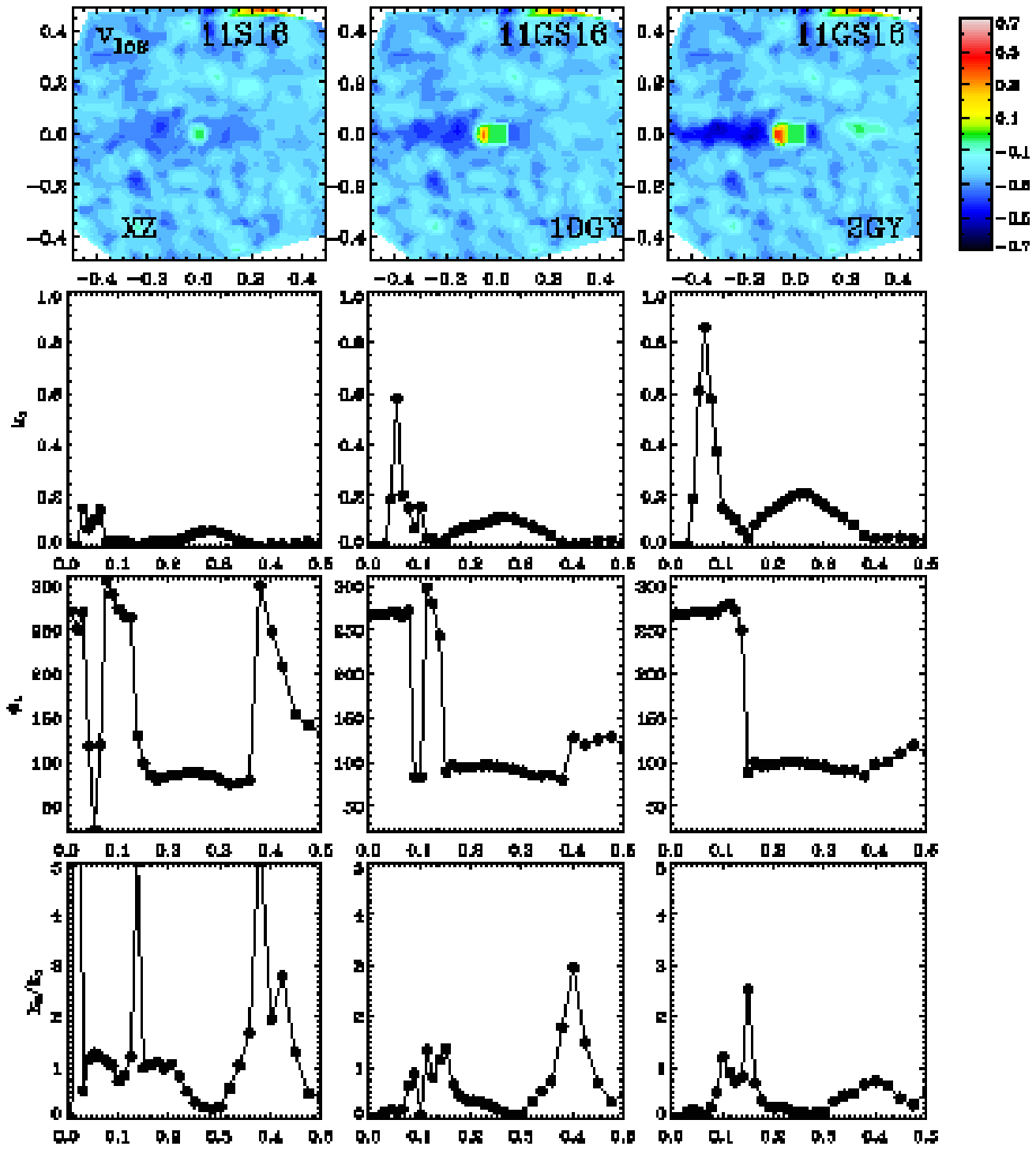, width=.8 \textwidth}
\caption{{\it Top row}: 2-dimensional velocity maps of the 11GS16 with a counter-rotating gas disk in the center.  
 The old stellar component {left}, as well as the gas and stars with the same M/L {middle} and lower M/L for the 
 gas particles {right} are shown. The kinemetry is extracted along ellipses with $q=0.3$ and $\Gamma=0^\circ$.
 {\it Second row}: The peak of the rotation curve ($k_1$) and the visibility of the CRC is improving from left to right.
  The CRC is hardly visible in the old stellar component. {\it Third row}: The $180^\circ$ kinematic position angle change is best visible
 in the 2Gy old population map. A second hump is visible at $r \approx 0.4$ and is a sign for the lopsidedness of the outer 
 gas disk. {\it Bottom row}: Again the 2Gy map shows the clearest transition signature in $k_5/k_1$.  }

\label{fig:2dcrcgas} 
\end{center}
\end{figure*}

Similarly we find kinematically misaligned gas, counter-rotating or otherwise, in 
about 50\% of the 1:1 merger remnants, but in only 1 out of 32 3:1 remnants. Our best example 
for a CRC in the {\it gas} is the merger remnant 11GS16, which originates from a very similar merging 
geometry ($i_1=60^\circ, \omega_1=90^\circ, i_2=-109^\circ, \omega_2=71^\circ$) than the original 
\citet{1991Natur.354..210H} calculation. In Fig. \ref{fig:2dcrcgas} we show edge on 2-dimensional velocity maps of 
this remnant. Kinemetry is uniquely suited to analyse maps with KDCs. K06 showed a very instructive example 
of a set of two-component test models in which they vary the alignment of the central component from aligned 
($\Delta \Gamma=0^\circ$) to counter-rotating ($\Delta \Gamma = 180^\circ$) (their Fig. 2) and proofed 
that the size and alignment of the KDC can be easily measured by kinemetry. We take advantage of this and 
extract the kinemetry of the stellar component of the gaseous remnant, of the gas amd stars with identical M/L
and of gas and stars with the gas particles having a lower M/L ('2Gy old population'). 

Observing the Gas-CRC at later or earlier times is crucial for the detectability in the maps. While kinemetry 
does pick up the signal in the '10Gy' observation (Fig.\ref{fig:2dcrcgas}, middle column), the '2Gy' map shows that
the possibility for an observer to  find this CRC would be much better. Interestingly the old stellar component 
is marginally dragged along into counter-rotation, although a weak signal is detectable. This CRC is relatively small,
about 10\% of the effective radius and the outer disk is lop-sided. 
The CRC is formed in this particular remnant due to a preservation of the original spin alingment 
of the progenitor discs. The rotation of particles which belonged to 
either disc 1 or disc 2 in the corresponding, i.e. originating from the same merging geometry, {\it collisionless} remnant 
are counter-rotating. Apparently in the dissipational remnant does the gas component of one progenitor galaxy 
lose more angular momnentum than the other and settle in a fast spinning disk in the center.

\begin{figure*}
\begin{center}
  \epsfig{file=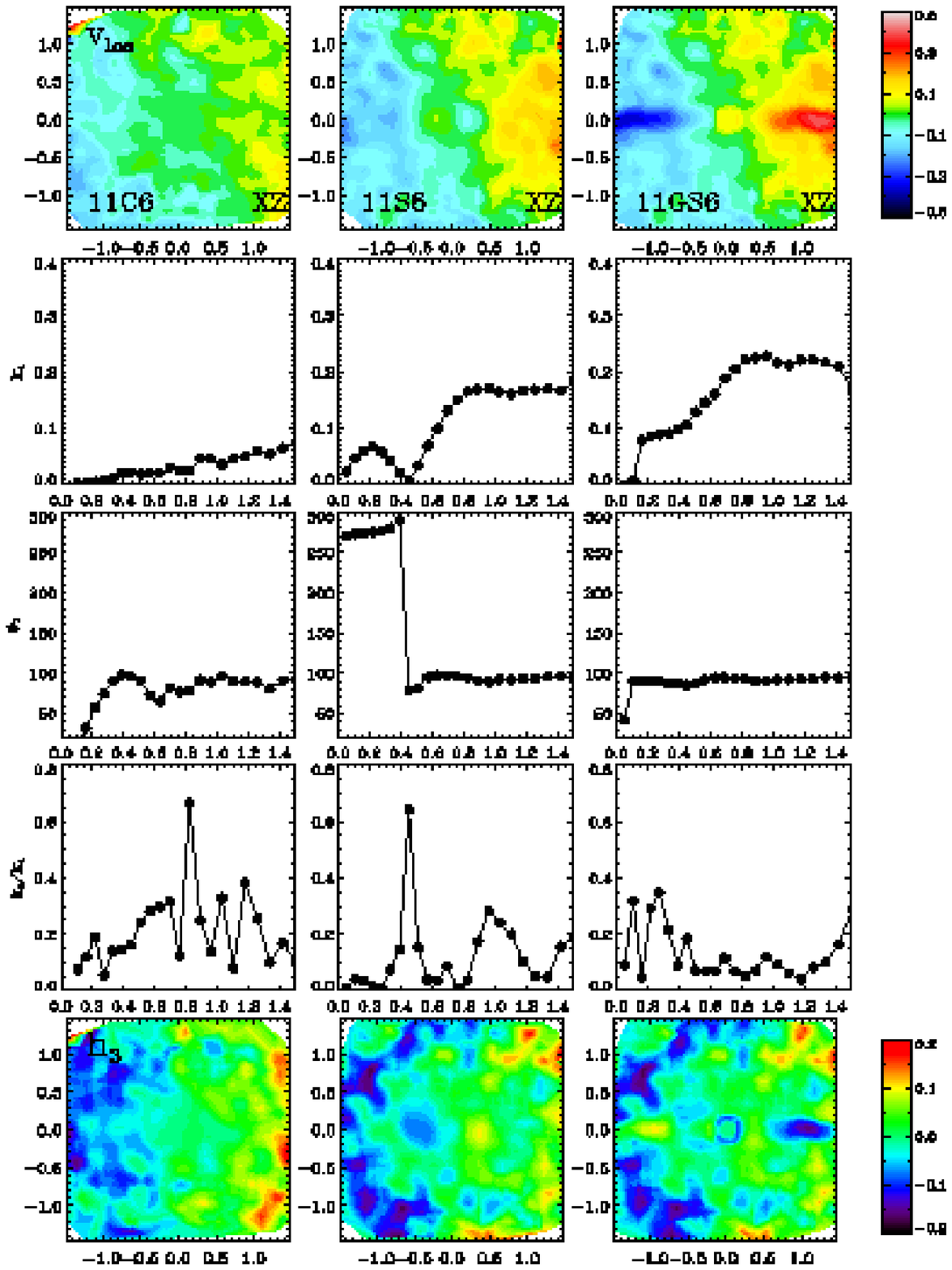, width=.8 \textwidth}
 \caption{{\it Top row}: Projected 2D line-of-sight
 velocity in the XZ projection for the collisionless remnant (11C6),
 the stellar remnant of the counterpart with gas (11S6) and 
 including the gas (11GS6) component assuming that
 all gas turned into stars after the merger was complete (from left to
 right). In the middle panel a counter-rotating stellar core (CRC) is
 clearly visible. In the right panel the kinematic signature of the
 thin disc appears. The ellipses, which have a $q=0.5$ and $\Gamma=0^\circ$,    
 along which we extracted the rotation curves are overplotted. 
{\it Second Row}: Rotation curves, $k_1$, extracted along an ellipse with $q=0.5$ 
and $\Gamma=0^\circ$ centered on the photometric major axis. The curves are shown
for the collisionless remnant (11C6), gas and stars of the dissipative remnant (11GS6) and
the stellar component of the gaseous remnant (11S6), which has a Counter-Rotating Core (CRC). 
{\it Third Row}: For 11S6, the ratio $k_5/k_1$ shows a strong peak at the transition radius between 
the CRC and the outer spheroid, as found in the test examples of K06. The other curves show 
no distinct features.
{\it Fourth Row}: Reconstructed angular direction $\phi_1$ of the rotation, $k_1$. 
11S6 shows a CRC with almost exactly $180^\circ$ misalignment in its rotation (i.e. counter-rotation). 
{\it Bottom row}: Two dimensional distribution of $h_3$ for the same remnants. Strong signatures for the
 counter-rotating core and the thin disc (middle and right panel).}
\label{fig:2dcrc} 
\end{center}
\end{figure*}

\begin{figure}
\begin{center} 
  \epsfig{file=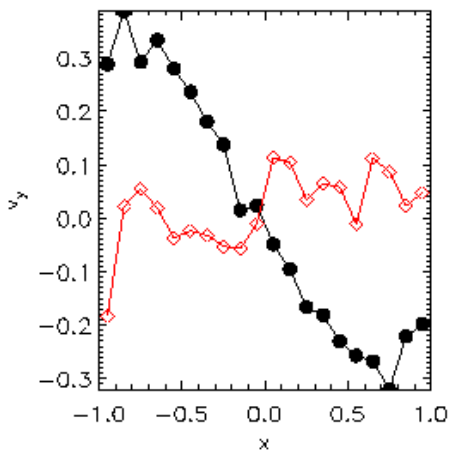, width=.8 \columnwidth}
\caption{ Line-of-sight velocity averaged in a thin slit along the major axis at the end of the simulation of remnant 11S6. 
The particles originally belonging to the disk of galaxy 1 (black filled circles) show a clear counter-rotation 
with respect to the particles orignally belonging to the bulge of the same galaxy (open diamonds).}
\label{fig:kdc_los} 
\end{center}
\end{figure}

\begin{figure*}
\begin{center} 
  \epsfig{file=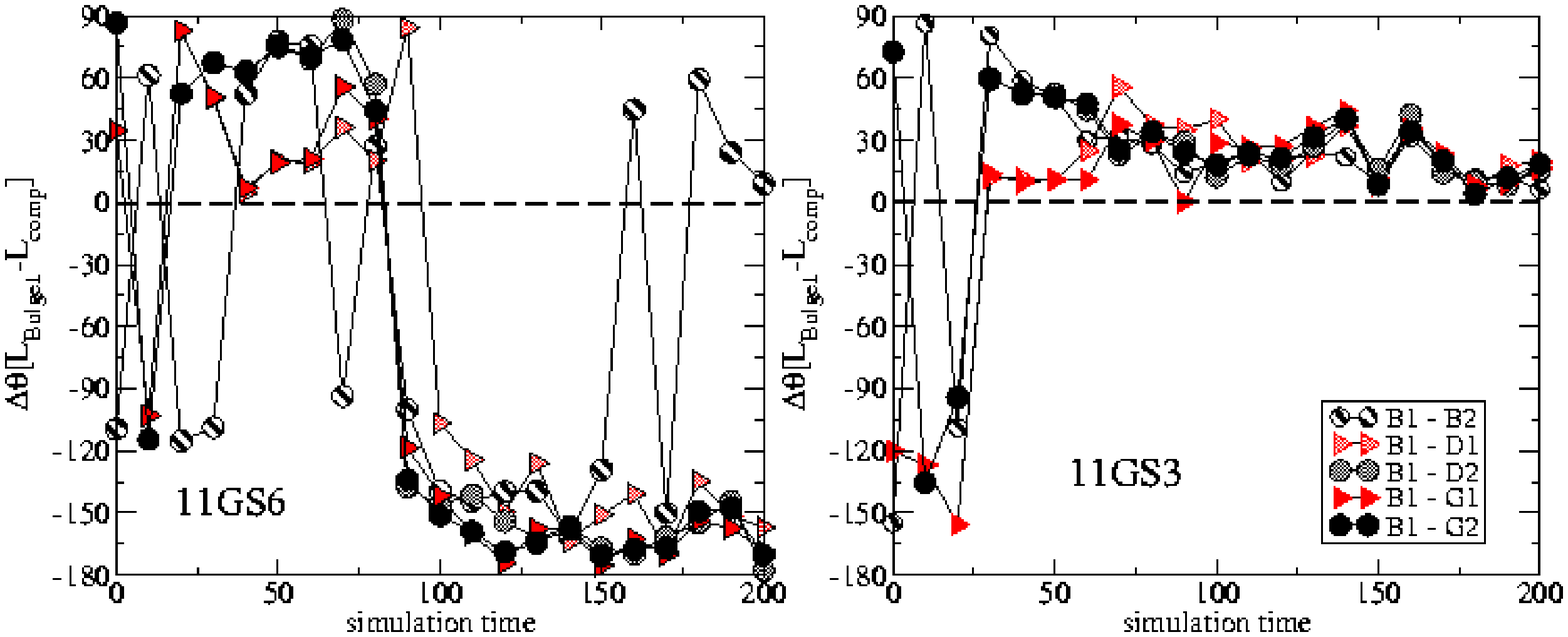, width=.9 \textwidth}
\caption{ {\it Left}: Evolution of the angle between the angular momentum vector of bulge1 particles of remnant 11S6 and all other
components: D1(disk1), B2(bulge2), D2(dik2), G1(gas disk1) and G2(gas disk2). Alignment is indicated by the thick horizontal black 
line at $0^\circ$. The spin of the Bulge1 is suddenly changed shortly after the second and final passage and stays at almost $-180^\circ$ with
respect to all other components. Only the particles of bulge 2 seem to evolve secularly and become more counter-rotating with time.
{\it Right}: The same properties plotted for remnant 11GS3, which has no CRC. The spin angles of all dynamical components 
are aligned shortly after the merging process. }
\label{fig:spin} 
\end{center}
\end{figure*}

However, the most clearly visible and largest CRCs are found in the old {\it stellar} component of equal-mass 
merger remnants. 11S6 is such an example and its maps and kinemetry is shown in Fig. \ref{fig:2dcrc}. We can neither 
detect a CRC in the collisionless remnant (top left), nor in the gas (top right), but only in the 'old stars' (top middle).
The CRC has a size of approximately half an effective radius and shows all tell-tale signs in the kinemetry, e.g. $180^\circ$ change 
of kinematic position angle. It is also visible in the $h_3$ showing an anti-correlation with $v_{los}$ as would be expected from a 
axisymmetric system with regular rotation. To study this phenomenon closer we examined the four distinct stellar 
components from which the particles which build  up the CRC can consist: disc of galaxy 1, bulge of galaxy 1, disk 
of galaxy 2 and bulge of galaxy 2. In Fig.\ref{fig:kdc_los} (top) we illustrate that the most strongly counter-rotating 
components are the particles from the disc and the bulge belonging orignally to the {\it same} galaxy (the particles of galaxy 2
lost most of their rotation and lie between those two extremes). We found it instructive to examine the evolution of the 
angle between the spin vectors of the bulge particles of galaxy 1 and the other three stellar components, repectively the
two other gas components. We calculate the angular momentum vector of each component with respect to its own
center of mass. In the same Fig. \ref{fig:spin}, left panel, we see that the alignment of the spin of the counter-rotating bulge
changes suddenly around $t=90$, which is shortly after the second, and final, encouter (around $t=70$). The bulge seems to be 
impulsively removed from its own disk. Later evolution shows that counter-rotation is slightly increasing and the angle difference 
stabilizes at $-150^\circ$ to $-180^\circ$. Only a significant part of the second bulge is at later times dragged 
into counter-rotation as well. This process is not occuring in all the remannts. The spin vector of all components, gas or stellar, of the 
remnant 11GS3 (Fig. \ref{fig:spin}, right panel) are aligned and no CRC is visible in the maps.

The formation of a CRC in such a context opens up questions if this is a robust result if realistic modes of 
star formation are included or if the subsequent merging history preserves such a KDC. Some aspects of forming 
a CRC in the old stellar component are favourable in light of recent observations and are discussed in Sec. \ref{sec:observ}.
These issues are beyond the scope of this paper and we try to address them in a subsequent paper.

\subsection{Line-of-sight velocity dispersion}

\begin{figure*}
\begin{center}
  \epsfig{file=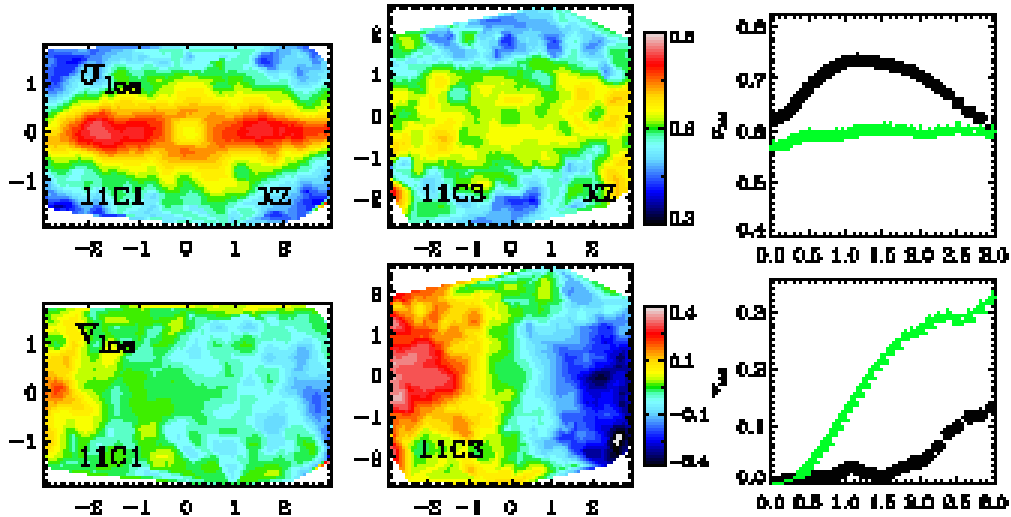, width=.8 \textwidth}
\caption{{\it Top row}: Velocity dispersion of two collisionless 1:1 merger remnants 11C1 and 11C3. The remnant 11C1 
has a high velocity dispersion disk-like component, as can be seen from the kinemetric profiles (black circles, right panel), 
significantly higher than 11C3 (green triangles). {\it Bottom row}: Because 11C1 originated from an in-plane 
merger of two disk with opposite spins is almost no net rotation visible. 11C3 shows regular rotation. }
\label{fig:2dsigmadisc} 
\end{center}
\end{figure*}

The  2-dimensional velocity dispersion maps of elliptical galaxies of the SAURON sample 
show a richness of features, which are even less understood than velocity maps in terms of their formation
history. Some of the galaxies show low dispersion regions, mostly in the center. Drops in $\sigma$ are indicative 
that dissipation has played a role in the formation of the galaxy.
It is indeed a very good indicator because lower velocity dispersion means that locally entropy has been taken 
out of the system and that would be hard to imagine in a pure collisionles process. However, collisionless
mergers can still produce non-trivial velocity dispersion maps as is shown in Fig. \ref{fig:2dsigmadisc}. 
The merger remnant 11C1 has a high $\sigma$ disk-like feature, which is a result of the peculiar merging geometry
($i_1=0^\circ, \omega_1=180^\circ, i_2=0^\circ, \omega_2=0^\circ$), i.e. both progenitors are in the same plane
and have anti-aligned angular momentum. The two superposed counter-rotating 'disks' in the final remnant maximize
the line-of-sight velocity dispersion. In contrast the merger 11C3 shows a more typical flat dispersion
profile. Both remnants have a dispersion dip in the center, but this signature is a left-over from the
cold centers of the initial Hernquist bulges \citep{1990ApJ...356..359H}. Also the velocity maps are quite 
different for both remnants. While the two counter-rotating components in 11C1 compensate each other, resulting
in almost zero $v_{los}$, does the remnant 11C3 show normal rotation. The high $\sigma$ disk phenomenon is probably rare,
 as 11C1 is the only remnant in our sample who has this feature, which is no surprise as the merging set-up needs to 
be rather fine-tuned.
 
\begin{figure*}
\begin{center}
  \epsfig{file=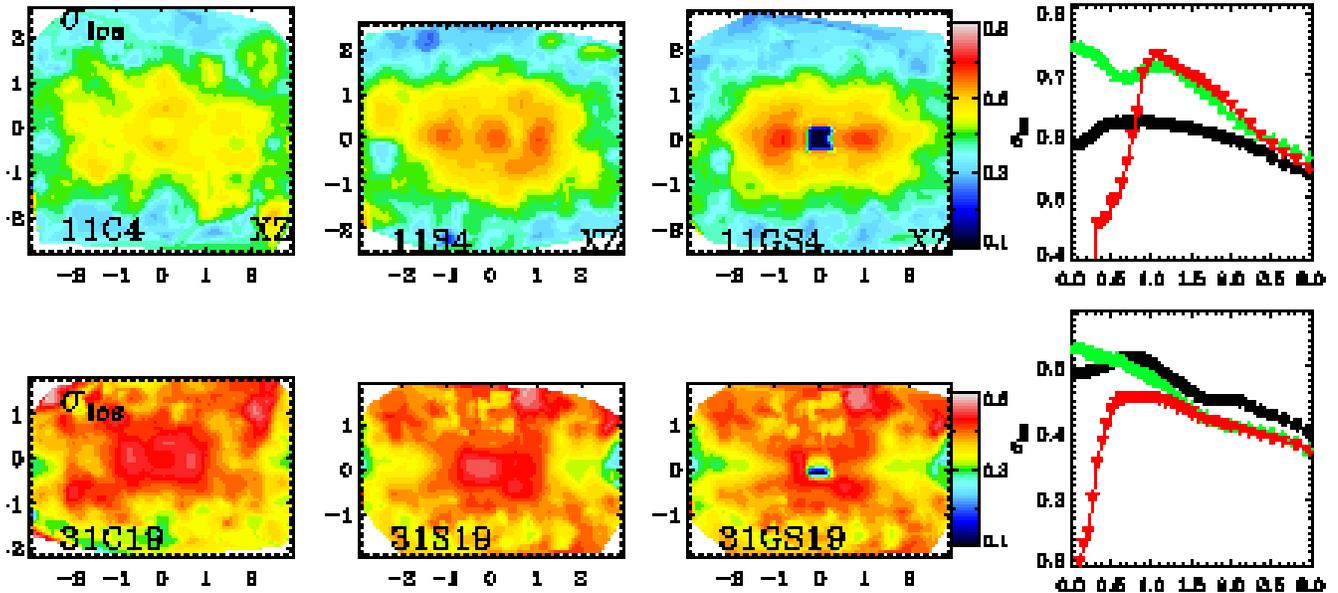, width=1. \textwidth}
  \caption{{\it Top row}: The velocity dispersion maps of the XZ projection for the 1:1 merger with 
  merging geometry 4. The velocity dispersion of the collisionless merger, 11C4, shows a slight
  depression of $\sigma_{los}$ at the center. The dispersion field of the stellar
  component of the gaseous remnant, 11S6, shows an off-axis double peak, which is even stronger if
  the gas particles are taken into account (11GS4). The kinemetry velocity dispersion extracted along an ellipse
  with $q=0.3$  is shown in the most right column for 11C4 (black circles), 11S4 (green triangles) and 11GS4 
  (red upside down triangles). Inclusion of the 
  {\it Bottom row}: 
  In the 31GS19 the gas settles partly into a cold stellar disc, which causes a $V$-shaped depression in the
  velocity dispersion.  }
\label{fig:2ddoublepeak}
\end{center}
\end{figure*}

Merger remnants with gas show a much wider variety in their velocity dispersion maps. We illustrate  
this with the example of a 1:1 and a 3:1 merger remnant (Fig. \ref{fig:2ddoublepeak}).
The dissipation  in the 3:1 remnant (bottom row) leads to a cold disk-like component, which imprints
a dumbbell like velocity dispersion structure. In general gaseous disks in un-equal mass mergers settle
into the equatorial plane, extend to larger radii and have more rotation than in equal-mass mergers.
The kinemetry shows that line-of-sight dispersion is lower than in the collisionless remnant at almost all 
radii (same figure, bottom right). Quite the contrary happens in the merger 11GS4, where a fraction of 
the gas particles accumulate into a high dispersion ring at about an effective radius from the center. 
Because the biggest fraction of the gas, like in all simulations, settles into a compact, dynamically cold, central 
disc, the dispersion map shows a double-peak structure (Fig. \ref{fig:2ddoublepeak}, top row). The peaks being 
edge-on cuts through the torus-like structure of the gas at this radius. The dissipation at this radius actually causes 
an increase the velocity dispersion with respect to the collisionless remnant (same figure, top right). 

\begin{figure*}
\begin{center}
  \epsfig{file=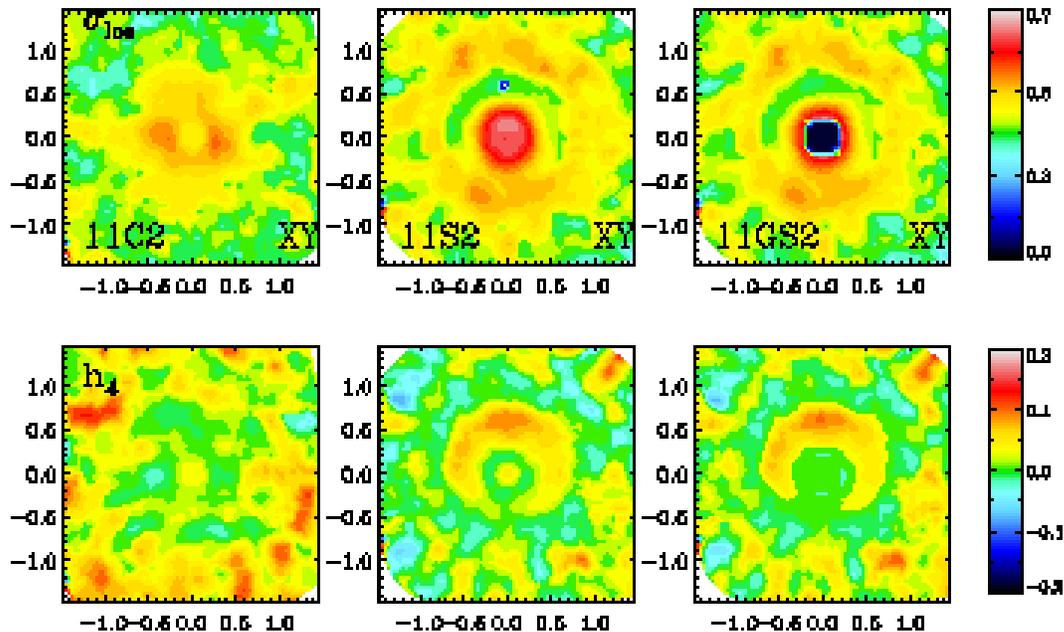, width=.8 \textwidth}
  \caption{Projected 2D velocity dispersion for remnant 11S2
  (top row) the the 2D distribution of $h_4$ for the
  same projections (bottom row). A depression of $\sigma$ at about
  $0.5r_e$ is clearly visible. In the same region $h_4$ changes sign 
  and becomes positive indicating a LOSVD more peaked than a Gaussian.}
\label{fig:2dsigmah4}
\end{center}
\end{figure*}

Also face-on (XY) views of gaseous remnants have distinctive features, such as ring-like depressions in the velocity
dispersion (top row of Fig. \ref{fig:2dsigmah4}). The most interesting aspect is that this feature is also imprinted
by the gas in the old stellar component. The low dispersion is not more pronounced If gas and star particles are observed 
simulataneously (11GS2), however the central gas component appears even colder in the map.   
The depression in $\sigma$ corresponds to a positive $h_4$ indicating a LOSVD that is more peaked than a 
Gaussian (same figure, bottom row).

\section{Comparison to Observations}
\label{sec:observ}
The previous sections highlighted the diversity of 2-dimensional maps of various moments of the 
LOSVD. We want to discuss in this section, how these results relate to observations of real early-type
galaxies, e.g. to the results of EM04. Unfortunately this comparison is mainly qualitative,
although we will loosely refer to the kinemetric categories as suggested by K06. For each category we 
will refer to the galaxies which are the principal examples for this category and comment on their
possible merger origin.

\noindent
{\bf Low-Rotation:} The kinemetry of the remnant 11C5 (Fig. \ref{fig:2drot}), shows 
(scaled) rotation of below 10 km/s inside one effective radius, which is extremely low-rotation. 
There are many galaxies in the SAURON sample that are not rotating outside a fast rotating 
central KDC, the so called slow rotators \citep{2005nngu.confE...5C}. 
We find three examples of galaxies with a rotation below 20 km/s: NGC 4374,  NGC 4486  and  
NGC 5846. All these objects are round and have elliptical or slightly boxy isophotes. Their high 
central velocity dispersion indicates that they are fairly massive and
that a merger of two very massive late-type galaxies would be needed to destroy the rotation in the 
progenitors and increase the mass accordingly. The scenario of merging elliptical galaxies would be even more 
favourable, as already the progenitors would have very little rotation \citep{2006ApJ...636L..81N}.

\noindent
{\bf Kinematic Twists and Misalignment:} A KT is indicated by a smooth change of the kinematic position angle. 
We find KTs only in 1:1 mergers which are violent enough to scatter stars from minor axis tubes 
onto major axis tube orbits. More than half of the 1:1 merger remnants show, at least at the center, iso-velocity contours twisting
more than $30^\circ$. The frequency and amplitude in the simulated equal-mass merger 
remnants seems to be higher than for the KTs found in the SAURON sample. In contrast, 3:1 mergers do not 
produce any galaxy with a twist of more than $20^\circ$, be it with or without gas. In this respect 
3:1 mergers are in better agreement with observations.

\noindent
{\bf Counter-Rotating Cores:} We define CRCs as a subgroup of KDCs with a sudden kinematic twist of
$180^\circ$. We find no CRCs in 3:1 remnants with or without gas and we find counter-rotating populations in two 1:1 
collisionless remnant, but which are not compact enough to pass as a core. For the stellar component of the 1:1 simulations 
with gas, we find at least four candidates for CRCs from the maps: 11S4, 11S6, 11S13 and 11S14 and one in the gas 11GS16. 
\citet{2006astro.ph..9452M} found in the SAURON galaxies two dinstinct types of KDCs: slow rotators harbour large 
KDCs (kpc scale) with ages and metallicities similar to the main body of the galaxy and fast rotators have smaller (below kpc scale)
and younger KDCs. Our simulations imply that stellar CRCs in slow-rotators are formed in 1:1 mergers when gas is 
present and, consistent with observations, as these CRCs consist of 'old' star particles in the simulation, they would 
not be detected as a distinct population, e.g. the CRC appears old. Additionally the main body would rotate slower, if he has
formed from a 1:1 merger. 
CRCs formed in the simulations in the gas are much smaller and younger, again in agreement with observations, but
we can hardly form them in 3:1 mergers. Most KDCs in fast rotators are actually CRCs, which might be a 
bias in the observations (see discussion  in \citealp{2006astro.ph..9452M}), still they seem too common place and merger 
simulations need to answer how to form them and maintain a high rotation in the main body.

\noindent
{\bf The $v-h_3$ Correlation:} Collisionless simulations show that there is generally  
a positive correlation between these quantities in the center of the remnants (NJB06). The presence of gas 
would inhibit the population of box orbits, the orbit class responsible for 
positive correlations (see discussion in Sec. \ref{sec:vh3}). Contrary to what is seen in the collisionless 
remnants, we see no correlation of $v-h_3$ in the center of any galaxy in the SAURON sample. In the outer parts there 
are a few examples with positive correlations: NGC 1023, NGC 2699, NGC 4270 and particularly NGC 4526. This could
be an indication for a presence of bars \citep{2004AJ....127.3192C}. This argues (see NJB06) against  pure collisionless 
mergers as possible formation mechanism of rotating SAURON galaxies, in agreement with the discussion on CRCs, see above.

\noindent
{\bf Opening Angle of Iso-Velocity Contours:} 
In Sec.  \ref{sec:qparam} we showed that on average 3:1 merger remnants with open iso-velocity contours have had less dissipation.
This is because dissipation causes flat discs and an overall more flattened and axisymmetric potential. 
As more material now rotates in the equatorial plane, the iso-velocity contours are closing resulting in a smaller $q$.
This result can not be easily compared in a statistical way to observations. Therefore we have chosen a pair of galaxies, which 
in the range of the SAURON observations have the same ellipticity $\epsilon$ 
and the same $v/\sigma$: these are NGC 7457 and NGC 4660, with $\epsilon=0.44$ and  $v/\sigma=0.6$ \citep{2005nngu.confE...5C}.
Although these galaxies appear in the $v/\sigma$-$\epsilon$ diagram at the same position, we find that the velocity 
maps of \citet{2004MNRAS.352..721E} of these two galaxies look very different: the velocity contours in 
NGC 7457 are much more open than those for NGC 4660. This indicates that much more dissipation was present in the formation 
of NGC 4660, probably causing a flattened disc-like component in this galaxy. The ellipticity in the outer parts, 
from the RC3 (de Vaucouleurs, 1993), also is very different: 0.41 for NGC 7457 and only 0.21 for NGC 4660. This gives support 
for the presence of a second flattened component in the middle of NGC 4660. 
For NGC 7457 we can not infer its intrinsic flattening, but it is likely to be constant with radius, since the ellipticity 
profile is constant \citep{1997NewA....1..349P}. 

\noindent
{\bf Polar Rings:} 
As laid out in Sec. \ref{sec:crc}, we find one very pronounced polar ring in the merger 11GS9. This polar ring is present in 
simulations with or without gas. In the SAURON survey the only two examples of polar rings are NGC 2685 and NGC 2768.
Both polar rings are only visible in the gas \citep{2006MNRAS.366.1151S}. The radial extent of the polar ring in the merger 
is very similar to what is found in these two observations. There appears to be a fundamental difference between simulations and
observations, as no stellar polar ring is found in the data. The statistics, however, is very small, so no strong conclusions 
can be drawn here. There are observational cases known of stellar polar rings (e.g. NGC 4365 \citealp{1995A&A...298..405S}; 
 \citealp{2001ApJ...548L..33D}), but these are generally small, not going as far out as one effective radius.

\noindent
{\bf Low $\sigma$ Rings:} In some equal-mass merger remnants with gas we find ring-like depressions in $\sigma$, 
mainly for face-on projections. Without gas we do not find them at all. Such a feature has been reported for  
one SAURON elliptical galaxy only: NGC 5813. In the spiral galaxy NGC 4314 \citep{2006astro.ph..3161F} a similar 
situation is visible: a star formation ring with a low gas velocity dispersion and a high stellar dispersion. 
In a recent survey of 6 nearby Seyfert galaxies \citep{2006MNRAS.371..170B} with IFU-GMOS, three of the galaxies, 
NGC2273, NGC3227 and NGC4593, show ring-like depressions in $\sigma$ in varying strength. If such low $\sigma$ rings 
are signatures of previous merging, it is interesting to note that most of them are found in active galaxies.

\noindent
{\bf Central $\sigma$ Drops:} In almost all merger remnants which were formed from progenitors with a dissipational
component, a significant fraction of the gas falls to the center, showing a very low $\sigma_{los}$ in the maps.
In contrast, the velocity dispersion of the stars is increasing towards the center. In general the SAURON observations 
for early-type galaxies also show peaks in $\sigma$ in the center, in agreement with the simulations. There are a few
objects with a central minimum in the velocity dispersion: these are NGC 4382 and NGC 2768. For spiral galaxies, 
recently \citet{2006MNRAS.tmp..171G} and \citet{2006astro.ph..3161F} found that dips in the central velocity dispersion 
are common, and that their frequency goes up with increasing Hubble type. They conclude that these dips are caused 
by central discs. In the case of the ellipticals the stars need about 1Gyr to
dynamically heat up after formation \citep{2003A&A...409..469W}.
So we suspect that in these two galaxies the velocity dispersion dips are caused by young stars. For NGC 4382 this could 
be consistent with the line-strength maps measured by \citep{2006astro.ph..2192K}, but this idea does not work for NGC 2768
for which old populations are inferred in that paper. In this galaxy there must have been a different mechanism 
to keep the stars in the central region cold. This would be consistent with the center of some Sa galaxies, where
velocity dispersion dips are found which are old (Peletier et al. in prep).

\noindent
{\bf Other $\sigma$ features:}
We found one example for a high $\sigma$ disk-like (11C1) structure caused by counter-rotation in the merger remnants. 
This feature is also rare in galaxies, NGC4473 being the only example in the SAURON survey (NGC4550 might 
also be comparable, but has a more overall complex structure). A merger origin seems very plausible for such a system 
although 11C1 has no net rotation and NGC4473 has still some. But the set-up of the merging discs gave them identical
amount of rotation with opposite sign which is very unlikely to happen in nature, such that some net rotation might well survive
a similar merging geometry. NGC2549 and especially NGC3384 have  dumbbell structures in their velocity dispersion maps and
NGC3377 has a strong low $\sigma$ disk. All these galaxies are fast-rotating, which fits well with our results, that we find such disks
predominantly in 3:1 mergers. However, we find them more abundantly than seen in EM04. The merger origin of NGC2685 has been
proposed before \citep{1999MNRAS.306..437H}. We add circumstantial evidence to this picture on the basis of the formation 
of a double-peaked velocity dispersion structure in some equal-mass mergers. The dissipational component and the violence
of the merging seem to be of the essence to puff-up stars and gas simultaneously.

In general, collisionless remnants show bigger opening angles of the iso-velocity contours, more boxy isophotes, 
velocity dispersion profiles without central peaks, no strong anti-correlation between $v-h_3$. 
Remnants with a dissipational component show smaller opening angles of the iso-velocity contours for 
the same ellipticities (inclinations), strong dips in the gaseous velocity dispersion profile and an anti-correlation 
between $v-h_3$. Additionally CRCs are more common in equal-mass mergers than in mergers of unequal mass. 
We conclude that globally the non-rotating subset of the representative SAURON sample of local galaxies agrees better
with our 1:1 merger simulations, while the rotating subset can be reproduced by the dissipational 3:1 merger remnants.

\section{Summary and Conclusions}
\label{sec:2dconcl}

We presented 2D maps of various moments of the 
LOSVDs of a large sample of 1:1 and 3:1 disc merger remnants with and without gas. 
Every remnant was resimulated with a dissipational component containing 10\% of the
luminous mass, allowing us to assess the influence of gas on the 2D
fields. Additionally we performed a kinemetric analysis using the method devised by K06 
to quantify properties such as kinematic position angle or deviations from regular 
rotation.

The difference between equal-mass and an unequal-mass merger remnants is not only seen in 
photometric and global kinematic properties \citep{1999ApJ...523L.133N}, but also in the 2D 
kinematics \citep{2000MNRAS.316..315B}, which exhibits very different features. 1:1 mergers can lead to a 
merger remnant with low rotation. Also, orbit classes rotating around the major axis of the remnant can 
only be populated in significant numbers in 1:1 mergers, leading to kinematic misalignment and twists, while 3:1 
remnants rotate much faster and have almost no kinematic twists. 
All rotating collisionless mergers fail to reproduce the observed $v-h_3$ anti-correlation, because the 
population of box orbits in the center is not inhibited (NJB06).

The presence of a dissipative component of only 10\% of the luminous disc mass in the progenitors
can lead to a considerable change in the properties of 1:1 remnants. They are more round, 
kinematic misalignment is reduced, counter-rotating cores are forming and kinematic twist are more pronounced. 
Rare but observed features like polar gas rings form also in equal mass mergers. The effect of gas 
on 3:1 remnants is less dramatic. KTs, CRCs and polar rings are not formed or at least must be very rare.  
A noticeable change is that the opening angles of the iso-velocity contours become smaller. 
As shown in NJB06, the $v-h_3$ anti-correlation is now much better reproduced, especially if the gas component 
is included in the analysis.

Merger remnants show a variety of velocity dispersion features. 
Significant central velocity dispersion dips are caused by infalling gas. The central star 
component in the gaseous runs, however, is heated. Edge on gas discs can exhibit lower
velocity dispersion (as found in 3:1 remnants) or edge-on gas rings can produce double-peaked velocity dispersion 
maps (as found in 1:1 remnants). Gaseous structures  can imprint a ring-like  depression in $\sigma$ even on 
the stellar component, which is also visible in $h_4$.

2D analysis of LOSVDs of N-Body simulations of galaxy formation  are an additional 
tool to connect the formation history of galaxies with their observable kinematic fine structure.
We defer a full statistical analysis to future work, with a larger sample of galaxies mergers, which
will also take star formation into account.

The simulations and analysis presented her are a step forward in understanding the possible formation 
of ellipticals by mergers of discs. We focus on the influence of a small dissipative component, which is 
important to understand its impact on the dynamical properties of the remnants. However, additional physical
processes like star-formation and feedback will change the detailed properties especially at the center 
of the remnants (\citealp{1996ApJ...464..641M};\citealp{2005ApJ...620L..79S}) and definitely have to be considered 
for simulations with higher gas fractions (\citealp{2004ApJ...606...32R};\citealp{2005ApJ...622L...9S}). 
We believe that the global effects of dissipation even with star formation will be very similar to the 
results presented here. However, larger gas fractions might result in more massive stellar discs in the remnant.
From a kinematic point of view disc merger remnants appear very similar to observed ellipticals, although
questions regarding the age and metalicity of the stellar populations have to be addressed in the future.
Analytical models of disc formation \citep{2006MNRAS.366..899N} as well as semi-analytical modeling \citep{2005astro.ph..9375K} 
in combination with merger simulations \citep{2006ApJ...636L..81N} can be used to place further constraints 
on the disc merger hypothesis.

\section*{Acknowledgments}
We are grateful to Davor Krajnovi\'c for kindly providing and helping with the implementation
of the Kinemetry software. The authors want to thank Karl Gebhardt 
for many helpful discussions and suggestions. This work was supported by the DFG 
priority programme 1177.
\bibliographystyle{mn2e}
\bibliography{references}

%\newpage
%\onecolumn
\appendix
\section{2D Velocity Maps and Kinemetric Data of Merger Remnants}
\label{sec:maps}
This is a selection of 2D velocity maps of all the remnants. We have to restrict ourselves
to the XZ projection, which is best suited for a kinemetric analysis, as it shows the highest 
rotational amplitude. For the equal-mass mergers we analyze the maps by extracting the rotation 
curve along circles, as the kinematic twists are too strong.
Although we do not study projection effects, we can see the merger to 
merger variance of the kinemetric parameters, such as the kinematic misalignment in 1:1 mergers. 
A full list of the orbital parameters for all merging geometries is given in Tab. \ref{tab:app}. 

\begin{table*}

\caption{Fulle List of merging geometries. For unequal-mass mergers the 
first number indicates  the orientation of the more massive 
galaxy as  $i_1$ and $\omega_1$,  the second number indicates the 
orientation of the more massive galaxy as  $i_2$ and $\omega_2$. \label{tab:app}}
\begin{tabular}{c|c|c|c|c}
\hline \hline
Geometry & $i_1$ & $i_2$  & $\omega_1$ & $\omega_2$ \\
\hline
1/17 & 0 & 0       & 180& 0   \\
2/18 & 0 & 0       & 71& 30   \\
3/19 & 0 & 0       & 71& -30  \\
4/20 & 0 & 0       & 71& 90   \\
5/21 & -109 & -60  & 180 & 0  \\
6/22 & -109 & -60  & 71 & 30  \\
7/23 & -109 & -60  & 71 & -30 \\
8/24 & -109 & -60  & 71 & 90  \\
9/25 & -109 & 0    & 180 & 0  \\
10/26 & -109& 0    & 71 & 30  \\
11/27 & -109 & 0   & 71 & -30 \\
12/28 & -109 & 0   & 71 & 90  \\
13/29 & -109 & 60  & 180 & 0  \\
14/30 & -109 & 60  & 71 & 30  \\
15/31 & -109 & 60  & 71 & -30 \\
16/32 & -109 & 60  & 71 & 90  \\
\hline
\end{tabular}
\end{table*}

%\vspace{15cm}
\begin{figure*}
\begin{center}
  \epsfig{file=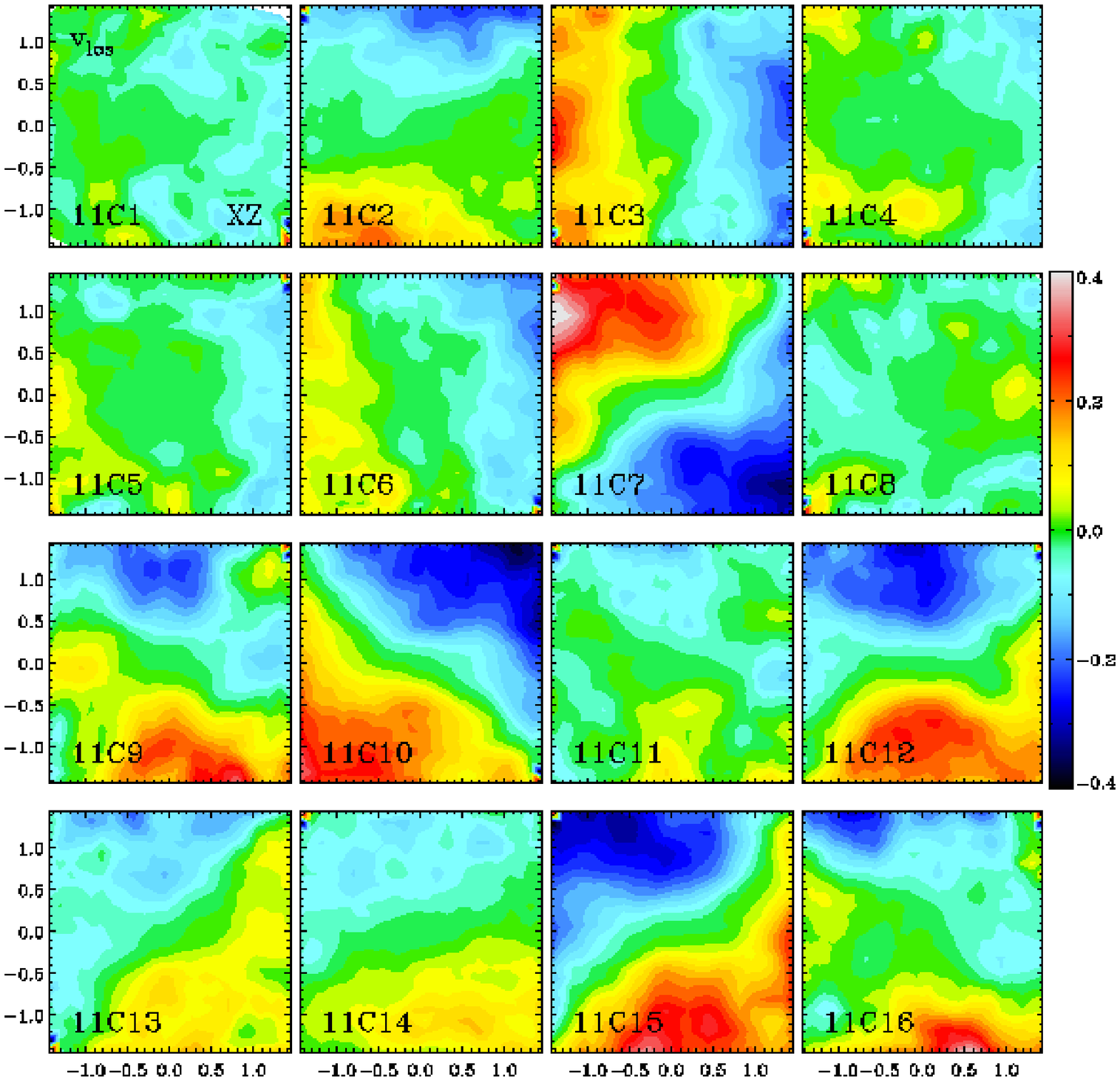, width=0.6 \textwidth}
  \epsfig{file=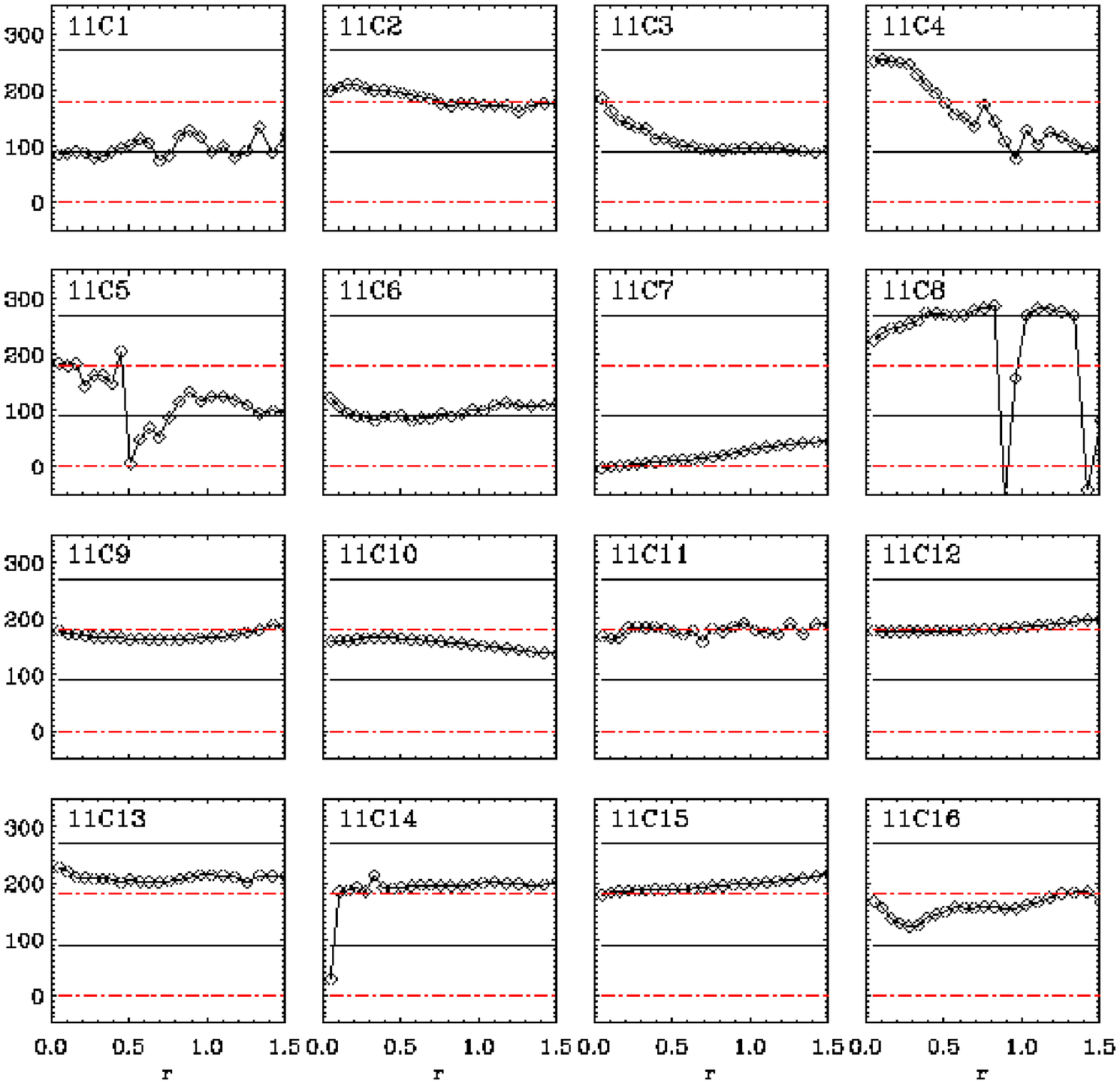, width=0.6 \textwidth}
 \caption{{\it Top:} Velocity maps for the collisionless 1:1 remnants for the XZ projection.
           {\it Bottom:} Corresponding kinematic position angles extracted along circles. By definition kinematic PAs 
          of $90^\circ$ and $270^\circ$ signify rotation which is aligned with the photometric major axis (indicated by straight 
          black lines), correspondingly $0^\circ$ and  $180^\circ$ indicate maximally misaligned rotation (red dot-dashed lines). 
          Strong kinematic misalignment is present in almost all remnants.} 
\label{VLOS_OVERVIEW_11C}
\end{center}
\end{figure*}

\begin{figure*}
\begin{center}
  \epsfig{file=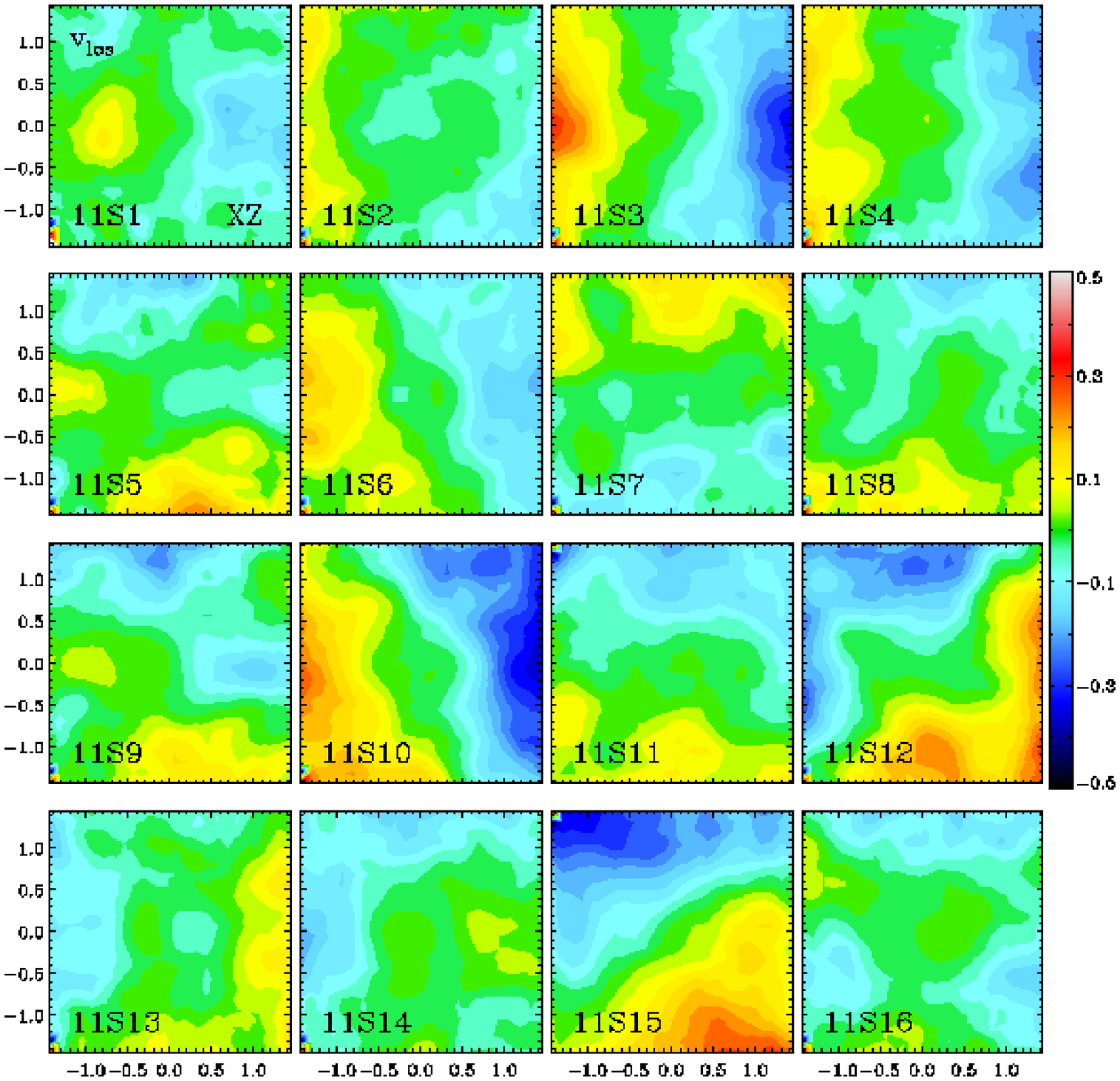, width=.6 \textwidth}
  \epsfig{file=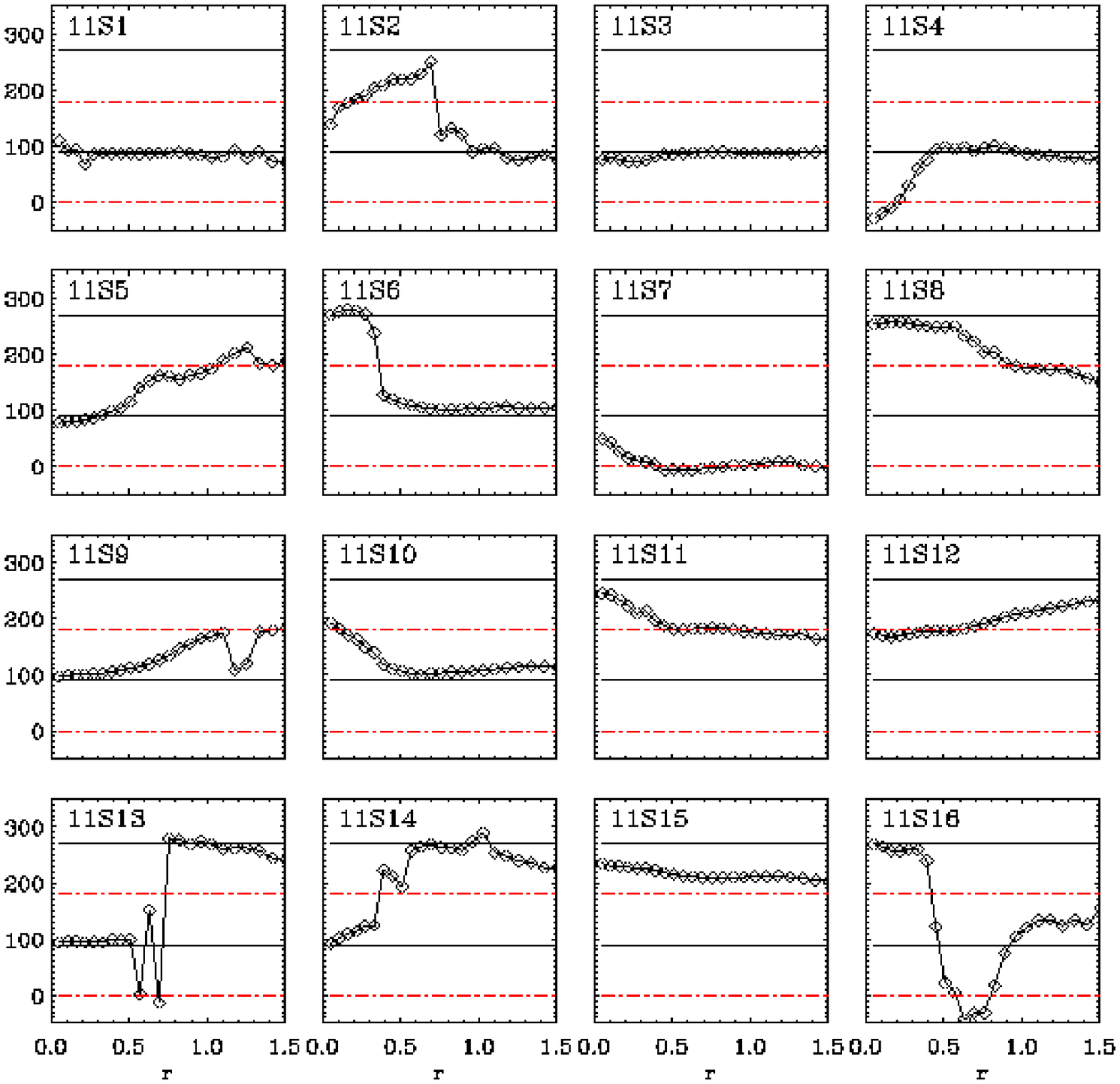, width=.6 \textwidth}
  \caption{{\it Top:} Velocity maps for the stellar component of the 1:1 remnants which 
           had progenitors with a gas component. {\it Bottom:} Corresponding kinematic position angles. Definition for horizontal lines 
as before. Sudden kinematic twists give indications for kinematic decoupled components in the center of the remnants. In the outer
parts the remnants are more aligned than their collisionless counterparts (see Fig. \ref{VLOS_OVERVIEW_11C})} 
\label{VLOS_OVERVIEW_11S}
 \end{center}
\end{figure*}
\begin{figure*}
\begin{center}
 will be available under http://www.usm.uni-muenchen.de
   \caption{{\it Top:} Velocity maps of the same 1:1 remnants like in the previous figure, 
         but with the gas particles converted into stars. {\it Bottom:} Kinemetric position angles as before. 
	Overall misalignment is reduced, some KDCs are not visible anymore, as the gas dominates the center.} 
\label{VKIN_OVERVIEW_11GS}
\end{center}
\end{figure*}

\begin{figure*}
\begin{center}
 will be available under http://www.usm.uni-muenchen.de
  \caption{The $k_1$ term (bulk rotation) and the $k_5/k_1$ (deviation from regular rotation) 
           are plotted with a commen x-axis. The low-rotation of 11C5, due to a high box orbit fraction in this remnant causes
            a high $k_5/k_1$ ratio.} 
\label{VKIN_KPARAM_11CS}
\end{center}
\end{figure*}

\begin{figure*}
\begin{center}
 will be available under http://www.usm.uni-muenchen.de
  \caption{The $k_1$ (bulk rotation) profiles and the $k_5/k_1$ ratio for the stellar component of remnants with gaseous 
	progenitors. Some CRCs are clearly visible in 11S6, 11S13, 11S14 and 11S16 in both a sudden peak 
        in $k_5/k_1$ and in the double hump structure of $k_1$.} 
\label{VKIN_KPARAM_11S}
\end{center}
\end{figure*}

\begin{figure*}
\begin{center}
will be available under http://www.usm.uni-muenchen.de
 \caption{Velocity maps for the collisionless 3:1 remnants for the XZ projection.} 
\label{VLOS_OVERVIEW_31C}
\end{center}
\end{figure*}

\begin{figure*}
\begin{center}

will be available under http://www.usm.uni-muenchen.de
 \caption{Velocity maps for the stellar component of the  3:1 remnants which had progenitors with a gas component} 
\label{VLOS_OVERVIEW_31S}
\end{center}
\end{figure*}

\begin{figure*}
\begin{center}

will be available under http://www.usm.uni-muenchen.de
 \caption{Velocity maps of the same 3:1 remnants like in the previous figure but with the gas particles converted into stars} 
\label{VLOS_OVERVIEW_31GS}
\end{center}
\end{figure*}

\end{document}